\documentclass[sigconf,screen]{acmart}

\AtBeginDocument{%
  }
    
\renewcommand\footnotetextcopyrightpermission[1]{}    

\usepackage{tcolorbox}
\usepackage{xspace}
\usepackage{enumitem}
\usepackage{subcaption}

\usepackage{amsthm}
\theoremstyle{acmdefinition}
\newtheorem{exmp}{Example}[section]

\usepackage{multirow}
\usepackage{tabularx}
\newcolumntype{C}[1]{>{\centering\arraybackslash}p{#1}}

\usepackage{microtype}
\usepackage{booktabs}

\newcommand{\todo}[1]{\textcolor{black}{#1}}
\newcommand{\todonew}[1]{\textcolor{black}{#1}}
\newcommand{\tool}{\textsc{LiVo}\xspace}

\newcommand{\modifyterm}{modification terms\xspace}
\newcommand{\MT}{MT}
\newcommand{\MTs}{MTs}

\begin{document}

\title{Detecting and Fixing Violations of Modification Terms in Open Source Licenses during Forking}

\author{Kaifeng Huang}
\affiliation{
\institution{Fudan University}
\city{Shanghai}
\country{China}
}

\author{Yingfeng Xia}
\affiliation{
\institution{Fudan University}
\city{Shanghai}
\country{China}
}

\author{Bihuan Chen}
\authornote{Bihuan Chen is the corresponding author.}
\affiliation{
\institution{Fudan University}
\city{Shanghai}
\country{China}
}

\author{Zhuotong Zhou}
\affiliation{
\institution{Fudan University}
\city{Shanghai}
\country{China}
}
\author{Jin Guo}
\affiliation{
\institution{Fudan University}
\city{Shanghai}
\country{China}
}

\author{Xin Peng}
\affiliation{
\institution{Fudan University}
\city{Shanghai}
\country{China}
}

\begin{abstract}
Open source software brings benefit to software community, but also introduces legal risks caused by license~violations, which~result in serious consequences such as lawsuits and financial losses.~To~mitigate legal risks, some approaches~have~been~proposed to identify licenses, detect license incompatibilities~and~inconsistencies, and recommend licenses. As far as we know,~however,~there~is no prior work~to understand modification terms in open~source~licenses or to detect and fix violations of modification terms.

To bridge this gap, we first empirically characterize modification terms in \todo{47} open source licenses. These licenses all require certain forms of ``notice'' to describe the modifications made~to~the~original work. Inspired by our study,~we~then~design~\tool~to~automatically detect and~fix~violations of modification terms in open~source~licenses during forking.~Our~evaluation~has~shown~the~effectiveness and efficiency of \tool.~\todo{18} pull requests of fixing modification~term violations have received positive responses. \todo{8} have been merged.
\end{abstract}

\maketitle



\section{Introduction}

Open source software has been widely embraced by developers, and software community has been fostered from~it~tremendously~in~the past decade. Despite the positive impetus~it~brings, open source software carries inherent risks, with one~of~the~most significant being~the~legal risks caused by violations~of~licenses. Open source licenses regard open source software as intellectual properties, and declare license terms that regulate the rights and obligations under which developers could conduct~activities~like reusing, distributing and modifying open source software.


There exists a large variety of open source licenses~\cite{opensourcelic},~ranging from permissive ones like MIT to restrictive ones like~GPL. Failing~to fulfill obligations in an open source license~can~cause a~license violation, also known as a license bug~\cite{vendome2018distribute}, which can hinder software development and prevent software from being released. One prominent example of license violations~is~the~enforcement~of copyleft terms in GPL. Copyleft requires that any modifications~to the software must also be distributed~under the same license as the original software, ensuring that the modified version remains open source like the original software. GPL~has enforced compliance on more than 150 products, resulting in many lawsuits and causing financial losses to companies~\cite{hemel2011finding}.


To mitigate legal risks, some approaches have been~proposed to identify licenses, detect license violations,~and~recommend~licenses. License identification is the first step towards~automatic~license~analysis. It is realized by rule-based~matching~\cite{tuunanen2009automated, german2010sentence, OSLC, gobeille2008fossology} and machine learning~\cite{higashi2016clustering, vendome2017machine, kapitsaki2017identifying}.~License~violations~can~be caused by various reasons.~One~line of work attempts~to~detect incompatibility among the licenses of a system and its declared dependencies or reused source code~\cite{kapitsaki2017automating, paschalides2016validate, xu2023lidetector}.~Another~line of work tries to find violations~of~copyleft terms of GPL/AGPL licenses~\cite{hemel2011finding, duan2017identifying, feng2019open}. The other line of work tries~to~detect~inconsistency between the licenses of two source files~that~are~evolved from the same provenance~by~code~reuse~\cite{german2009code, wu2015method, wu2017analysis}~and~between the license of a package and the license~of~each~file~of the package~\cite{german2010understanding, di2010identifying, mlouki2016detection}. To prevent license violations, license~recommendation~\cite{liu2019predicting, kapitsaki2019modeling} can be used to choose compliant licenses.

However, little attention has been paid to violations~of~modification terms (MTs) in open source licenses. MTs specify~the conditions under which users of open source software~may~modify the original software~and the obligations~they~must~fulfill. For example, the Apache-2.0 license specifies that \textit{``You~may~reproduce and distribute copies of the Work ..., provided~that~you meet the following conditions: ... b. You must cause~any~modified files to carry prominent notices stating that You changed~the files; ...''}. MTs distinguish liability, copyright and benefit~among shareholders when distributing derivative works. For example, the NGPL license~dictates~that~\textit{``If~NetHack is modified by someone else and passed on, we want its recipients to know that what they have is not what we distributed''}. \todonew{Besides, developers may also not be familiar with MTs, and thus discuss how to obligate MTs~\cite{howtospecify, filewithapache, tmvdiscussion}. Therefore, it is an issue worthy of investigation.} Unfortunately,~to~the~best of our knowledge, there has been no effort to understand MTs~in open source licenses or to detect and fix their violations in an automated way.





To fill this gap, we first conduct an empirical~study~on~MTs~of open source licenses approved by Open~Source~Initiative~\cite{opensourcelic}. Of these \todo{107} approved licenses, \todo{47} licenses noticeably~declare MTs, and require certain forms~of ``notice'' to describe~modifications made to the original~work. To~gain~a deeper understanding~of~MTs, we analyze~each license document to identify similarities and differences~in~how MTs are implemented. We find that they differ in modification scope (i.e., the~types~of modified files), notice content (i.e., the description of modifications), and notice location (i.e., the location to put the notice). We summarize \todo{eight}, \todo{six} and \todo{four} categories of modification scope, notice content and notice location, respectively, which are used to model MTs of the \todo{47} studied licenses.


Inspired by our study, we then propose~an~automatic~approach, named \tool, to detect and fix violations~of MTs~during forking, where a fork repository~is~first~created~by copying~a base repository and then modified separately.~We~specifically focus~on~the~reuse and redistribution scenario of forking because i) it is widely used~in open source software development,~and ii) it is non-trivial to ensure compliance with~MTs~due~to~the~complex modification history of the fork repository.~Given~a~base and a fork repository,~\tool~works~in three steps. First,~it~identifies the obligating commits~of~the~fork~repository. These~obligating commits modify the original files that~are~in the modification scope of the MT of the base repository. Second,~it~extracts change logs from potential notice location in the fork repository. These change logs~contain potential~notice for those obligating commits. Third, it matches change logs with commit logs based on notice content of the MT to detect and fix MT violations. \todo{\tool can be adopted by authors of the base repositories to detect MT violations, and by authors of the fork repositories to detect and fix MT voliations such that potential legal risks can be avoided.}

We evaluate the effectiveness and efficiency of \tool~using \todo{178} pairs of base and fork repositories. \todo{91} fork repositories violate MTs of the base repositories, while \todo{87} fork repositories do not~modify original files and hence fulfill MTs of the~base repositories. 
\tool achieves a precision of \todo{0.82} and~a~recall~of \todo{0.80} in detecting~and~fixing MT violations. We~submit \todo{91} pull requests to report~and~fix MT~violations. \todo{18}~of them have received positive responses from developers, and \todo{8}~have been merged. \tool takes \todo{229.8} seconds to detect and fix MT violations in each pair of base and fork~repository.

In summary, our work makes the following contributions.
\begin{itemize}[leftmargin=*]
    \item We conducted the first empirical study on \todo{47} open source licenses to characterize MTs.
    \item We proposed \tool to detect and fix violations~of~MTs in open source licenses during forking.
    \item We conducted evaluation to demonstrate the effectiveness and efficiency of \tool. \todo{18} pull requests of fixing MT violations have received positive responses, and \todo{8} have been merged.
\end{itemize}




\section{Understanding Modification Terms}\label{sec:understanding}

\todonew{We design an empirical study to understand modification terms~(MTs) in open source licenses by answering four research questions.}

\begin{itemize}[leftmargin=*]
    \item \todonew{\textbf{RQ1 Modification Scope Analysis:} How do MTs define the scope of modification to the original work?}
    \item \todonew{\textbf{RQ2 Notice Content Analysis:} What are the required contents to put into the notice for obligating MTs?}
    \item \todonew{\textbf{RQ3 Notice Location Analysis:} How do MTs define the location where the required notice should be put?}
    \item \todonew{\textbf{RQ4 Obligation Analysis:} What is the obligation of MTs in open source licenses?}
\end{itemize}

\todonew{To answer these research questions, we collect open source li- censes that declare MTs, and label each MT with respect to modification scope, notice content and notice location. Then, we model obligations of MTs with the above three dimensions.}


\subsection{Collecting and Labeling MTs}

To understand MTs in open source licenses, we first collect a total of \todo{107} approved licenses which are listed on Open~Source Initiative (OSI)~\cite{opensourcelic}. OSI is a non-profit~corporation~actively~involving in~and advocating open source. These licenses include open source licenses such as Apache-2.0, BSD and MIT. Then, two of the authors read each license document~to~find~MTs~and obligations in the MTs. They reach a Cohen's Kappa coefficient of \todo{0.887}, and conflicts are resolved by involving a third author. 



\todo{10} of the \todo{107} open source licenses~do~not mention modifications. \todo{40} open source licenses mention~modifications,~but~do not require any obligation on modifications. For example, the~MIT license does not require further~obligation on modifications. It dictates that \textit{``Permission is hereby granted, free of charge, ... to deal~in~the Software without restriction, including without limitation the rights~to use, copy, \textbf{modify}, merge, publish, ...''.} \todo{47} open source licenses declare MTs, and require certain~forms of ``notice'' to describe~modifications made to the original~work as the obligation. \todo{10} open source licenses mention modifications, and require other forms of obligation (e.g., putting trademarks).


To characterize MTs in these \todo{47} licenses, two~of~the~authors discuss obligations of MTs \todo{and extensively refer to online third-party resources} after reading each MT,~\todo{until reaching}~an agreement on modeling the obligation~of~a~MT~from three dimension, i.e., modification scope, notice content and notice location.~To label each MT with respect to these three dimensions, two of the~authors follow an open coding procedure \cite{seaman1999qualitative} to inductively create composing elements and categories for the three dimensions~by~analyzing each MT. We use Cohen’s Kappa coefficient to measure~agreement, and it reached \todo{0.760}, \todo{0.731} and \todo{0.813}, respectively.

\begin{table}
    \centering
    \scriptsize
    \caption{Composing Elements of Modification Scope, Notice Content, and Notice Location}
    \vspace{-10pt}
    \begin{subtable}[!t]{0.45\linewidth}
        \centering
        \caption{\todo{RQ1:} Modification Scope}\label{table:scope_components}
        \vspace{-6pt}
        \(\begin{array}{m{0.2cm}m{3.8cm}}
            \noalign{\hrule height 1pt}
            ID       & Composing Element                           \\
            \noalign{\hrule height 1pt}
            $S_{c_1}$ & Source Code                         \\\hline
            $S_{c_2}$ & Documentation                       \\\hline
            $S_{c_3}$ & Configuration Files                 \\\hline
            $S_{c_4}$ & Interface Definition Files          \\\hline
            $S_{c_5}$ & Scripts                             \\\hline
            $S_{c_6}$ & Source Code Differential Comparison \\\hline
            $S_{c_7}$ & Design Materials                    \\\hline
            $S_{c_8}$ & Others                              \\
            \noalign{\hrule height 1pt}
        \end{array}\)
    \end{subtable}
    \hfill
    \begin{subtable}[!t]{0.4\linewidth}
        \caption{\todo{RQ2:} Notice Content}\label{table:content_component}
        \vspace{-6pt}
        \(\begin{array}{m{0.2cm}m{2.3cm}}
            \noalign{\hrule height 1pt}
            ID       & Composing Element             \\
            \noalign{\hrule height 1pt}
            $C_{c_1}$ & Date                  \\\hline
            $C_{c_2}$ & Author            \\\hline
            $C_{c_3}$ & Brief Statement       \\\hline
            $C_{c_4}$ & Informative Statement \\
            \noalign{\hrule height 1pt}
        \end{array}\)
        \medskip
        \vspace{-4pt}
        \caption{\todo{RQ3:} Notice Location}\label{table:location_component}
       \vspace{-7pt}
        \(\begin{array}{m{0.2cm}m{2.3cm}}
            \noalign{\hrule height 1pt}
            ID       & Composing Element          \\
            \noalign{\hrule height 1pt}
            $L_{c_1}$ & A Document         \\\hline
            $L_{c_2}$ & Each Modified File \\
            \noalign{\hrule height 1pt}
        \end{array}\)
    \end{subtable}
\end{table}

\begin{table*}[!t]
    \centering
    \scriptsize
    \setlength{\leftmargini}{0.4cm}
    \caption{\todo{RQ1:} Groups of Modification Scope }\label{table:scope_groups}
    \vspace{-10pt}
    \begin{tabular}{m{0.5cm}m{14cm}m{2.2cm}}
        \noalign{\hrule height 1pt}
        Group    & \MT~Excerpted from License Document                                                                                                                                                                                                                                                                                                                                                                                                                         & Composing Elements                                        \\\hline
        \noalign{\hrule height 1pt}
        $S_{g_1}$ & \textbf{AFL-3.0}: The term ``Source Code'' means the preferred form of the Original Work for making modifications to it and all \textbf{available documentation describing how to modify the Original Work}.                                                                                                                                                                                                                                                  & $S_{c_1}$, $S_{c_2}$                               \\\hline
        $S_{g_2}$ & \textbf{OHL-2.0}: `Source' means information such as \textbf{design materials or digital code} which can be applied to Make or test a Product or to prepare a Product for use, Conveyance or sale, regardless of its medium or how it is expressed. It may include Notices.                                                                                                                                                                                 & $S_{c_1}$, $S_{c_7}$                               \\\hline
        $S_{g_3}$ & \textbf{Apache-2.0}: ``Source'' form shall mean the preferred form for making modifications, including but not limited to \textbf{software source code}, \textbf{documentation source}, and \textbf{configuration files}.                                                                                                                                                                                                                                     & $S_{c_1}$, $S_{c_2}$, $S_{c_3}$                     \\\hline
        $S_{g_4}$ & \textbf{APSL-2.0}: ``Source Code'' means the human readable form of a program or other work that is suitable for making modifications to it, including all modules it contains, plus \textbf{any associated interface definition files, scripts} used to control compilation and installation of an executable (object code).                                                                                                                                 & $S_{c_1}$, $S_{c_4}$, $S_{c_5}$                     \\\hline
        $S_{g_5}$ & \textbf{MPL-1.1}: ``Source Code'' means the preferred form of the Covered Code for making modifications to it, including all modules it contains, plus any \textbf{associated interface definition files, scripts} used to control compilation and installation of an Executable, or \textbf{source code differential comparisons} against either the Original Code or another well known, available Covered Code of the Contributor's choice.              & $S_{c_1}$, $S_{c_4}$, $S_{c_5}$, $S_{c_6}$           \\\hline
        $S_{g_6}$ & \textbf{RPL-1.5}: ``Source Code'' means the preferred form for making modifications to the Licensed Software and/or Your Extensions, including all modules contained therein, plus any \textbf{associated text, interface definition files, scripts} used to control compilation and installation of an executable program ...                                                                                                                                & $S_{c_1}$, $S_{c_2}$, $S_{c_4}$, $S_{c_5}$           \\\hline
        $S_{g_7}$ & \textbf{SPL-1.0}: ``Source Code'' means the preferred form of the Covered Code for making modifications to it, including all modules it contains, plus any \textbf{associated documentation, interface definition files, scripts} used to control compilation and installation of an Executable, or \textbf{source code differential comparisons} against either the Original Code or another well known, available Covered Code of the Contributor's choice. & $S_{c_1}$, $S_{c_2}$, $S_{c_4}$, $S_{c_5}$, $S_{c_6}$ \\\hline
        $S_{g_8}$ & \textbf{GPL-3.0}: To ``modify'' a work means to copy from or adapt \textbf{all or part of the work} in a fashion requiring copyright permission, other than the making of an exact copy. The resulting work is called a ``modified version'' of the earlier work or a work ``based on'' the earlier work.                                                                                                                                                         & $S_{c_1}$--$S_{c_8}$                                \\
        \noalign{\hrule height 1pt}
    \end{tabular}
\end{table*}

\begin{figure*}[!t]
    \centering
    \begin{subfigure}[b]{0.24\textwidth}
        \centering
        \includegraphics[width=0.99\textwidth]{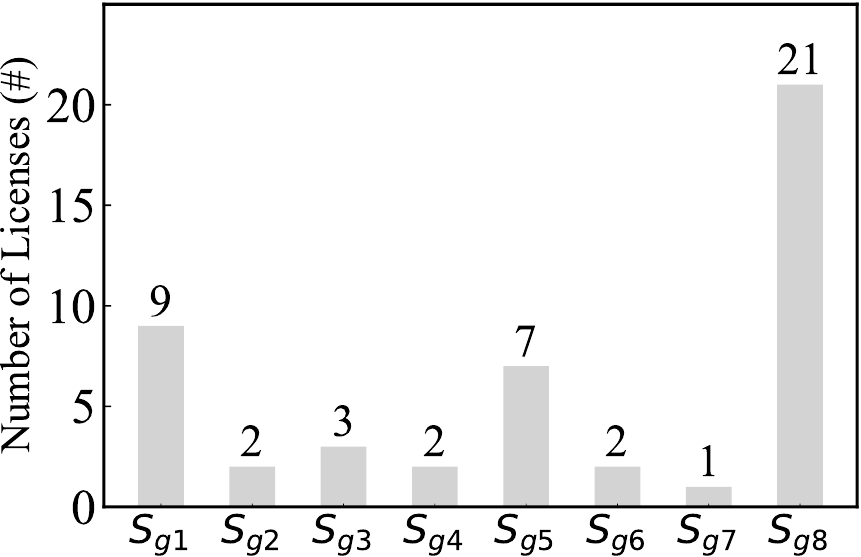}
        \vspace{-10pt}
        \caption{\todo{RQ1:} Modification Scope}\label{fig:scope}
    \end{subfigure}
    \begin{subfigure}[b]{0.24\textwidth}
        \centering
        \includegraphics[width=0.99\textwidth]{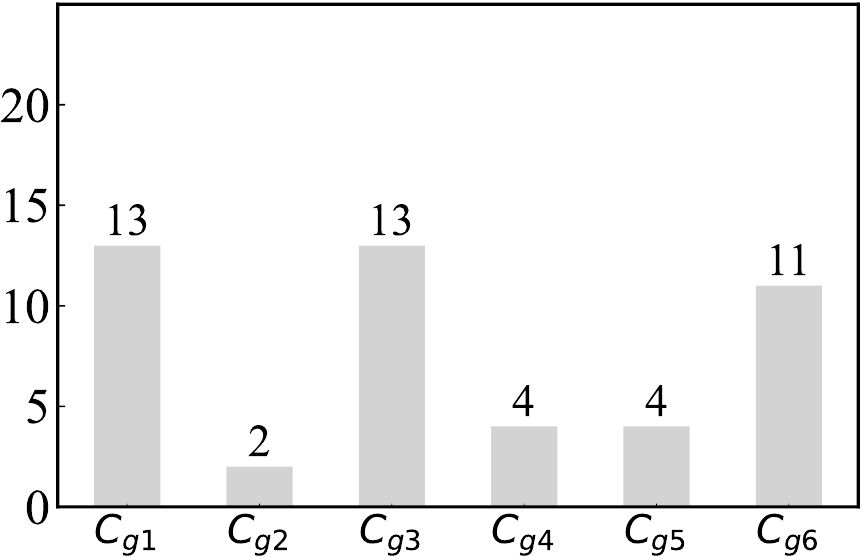}
        \vspace{-10pt}
        \caption{\todo{RQ2:} Notice Content}
        \label{fig:content}
    \end{subfigure}
    \begin{subfigure}[b]{0.24\textwidth}
        \centering
        \includegraphics[width=0.99\textwidth]{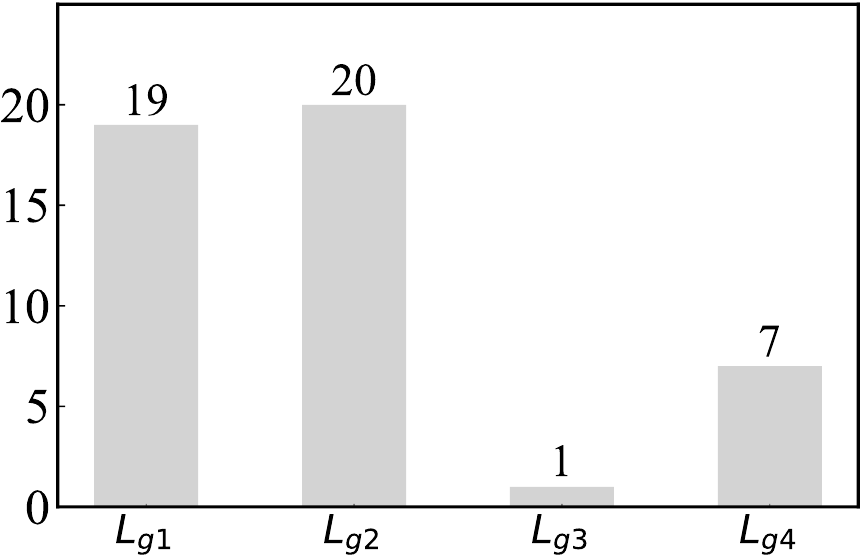}
        \vspace{-10pt}
        \caption{\todo{RQ3:} Notice Location}\label{fig:location}
    \end{subfigure}
    \begin{subfigure}[b]{0.24\textwidth}
        \centering
        \includegraphics[width=0.99\textwidth]{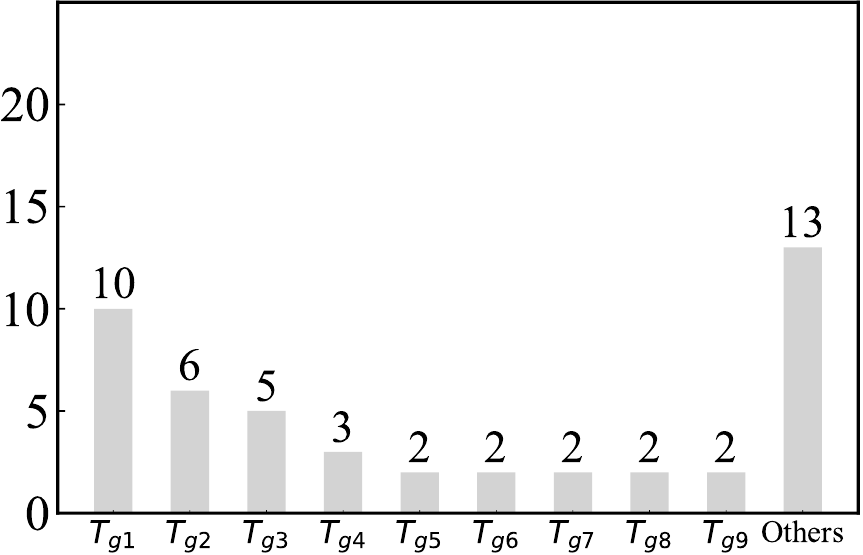}
        \vspace{-15pt}
        \caption{\todo{RQ4:} Obligation Group}\label{fig:term_group}
    \end{subfigure}
    \vspace{-6pt}
    \caption{License Distribution across Groups of Modification Scope, Notice Content, Notice Location and Obligation Group}\label{fig:obligation}
\end{figure*}

\subsection{Modeling Obligations of MTs}\label{sec:mt-modeling}

We model obligations of MTs with modification scope,~notice content, and notice location. Specifically, modification~scope~describes the types of original files that~are~modified~(e.g.,~source code and documentation). Notice content represents the description of modifications required by the obligation~of~a~MT.~It~is in the~form of natural language statements with some meta-data (e.g., author). Notice location represents the location~to~put the notice~content~required by the obligation (e.g., a document).~If users of an open~source software modify the original files~that are in the modification scope of a MT, they must put required notice content to a required notice location~to~fulfill~the~obligation. Otherwise, the MT is violated. Open source licenses differ significantly in the obligation of MTs. We~summarize and introduce composing elements and categories of each dimension \todo{(i.e., \textbf{RQ1}, \textbf{RQ2} and \textbf{RQ3})}, and then present the obligation modeling based on these three dimensions \todo{(i.e., \textbf{RQ4})}.


\textbf{Modification Scope (\todo{RQ1}).} We summarize \todo{eight} composing elements~of modification scope (denoted as $S_{c_i}$, $1 \le i \le 8$), i.e., \textit{Source Code},~\textit{Documentation}, \textit{Configuration Files}, \textit{Interface Definition Files}, \textit{Scripts},~\textit{Source Code Differential Comparison}, \textit{Design Materials}, and \textit{Others}, as shown in Table~\ref{table:scope_components}. The file types of $S_{c_1}$ to $S_{c_7}$ are easy to understand.~We~add \textit{Others} ($S_{c_8}$) to denote the rest of file types not listed in $S_{c1}$~to~$S_{c7}$~because some licenses require obligation on modifications to all files.



Based on these composing elements, we categorize licenses into \todo{eight} groups (denoted as $S_{g_i}$, $1 \le i \le 8$) with respect to the modification scope of their MTs, as presented in Table~\ref{table:scope_groups}. The second~column illustrates~an~example of MT partially excerpted from~a~license~that belongs to each group. The third column reports the specific composing elements of the modification scope for each group. Notice that we use $S_{c_1}$--$S_{c_8}$ to denote all files from the original work. 
\textit{Source Code} ($S_{c_1}$) is included in the modification scope of~all~MTs.


\begin{exmp}\label{exmp:scope_group}
The MT of Apache-2.0 dictates that \textit{```Source' form~shall mean the preferred form for making modifications, including~but not limited to software source code, documentation source, and configuration files''}. Thus, it belongs to $S_{g_3}$ whose modification~scope~includes $S_{c_1}$, $S_{c_2}$ and $S_{c_3}$. The MT of GPL-3.0 declares that \textit{``To `modify'~a~work means to copy from or adapt all or part of the work''}. Thus, we put it into~$S_{g_8}$ whose modification scope includes all files, i.e., $S_{c_1}$ to $S_{c_8}$.
\end{exmp}

We analyze the distribution of \todo{47} open source licenses~across the \todo{eight} groups of modification scope. The result is reported~in Fig.~\ref{fig:scope}. \todo{21} (\todo{44.7\%}) licenses belong~to~$S_{g_8}$, accounting for the largest portion of licenses. This indicates that nearly half~of~the licenses require obligation on modification~to~any~file.~\todo{9}~(\todo{19.1\%}) and \todo{7} (\todo{14.9\%}) licenses respectively belong to $S_{g_1}$ and $S_{g_5}$.~Each of the~rest five groups contains no more than three licenses. 

\textbf{Notice Content (\todo{RQ2}).} We summarize \todo{four} composing elements~of notice content (which are denoted as $C_{c_i}$, $1 \le i \le 4$), i.e., \textit{Date}, \textit{Author}, \textit{Brief Statement} and \textit{Informative Statement}, as listed in Table~\ref{table:content_component}.~\textit{Date} and \textit{Author} describe when and who modifies~the original files. We distinguish modification description into \textit{Brief Statement}~and \textit{Informative Statement}. The former only requires to state that you change the files, while the latter also requires to state that how you change the files, the detail~of~the~change,~etc.


Based on these composing elements, we categorize licenses into \todo{six} groups (denoted as $C_{g_i}$, $1 \le i \le 6$) with respect to the notice~content of their MTs, as shown in Table~\ref{table:content_groups}. Similar~to Table~\ref{table:scope_groups},~the~second column shows~an~example of MT,~and~the third column lists the specific composing elements.


\begin{exmp}\label{exmp:content_group}
The MT of LGPL-2.1 declares that ``\textit{You must cause the files modified to carry prominent notices~stating~that you changed~the files and the date of any change}''.~It~belongs to $C_{g_1}$~as~it~requires~to~include \textit{Date} and \textit{Brief Statement} into notice~content.~The~MT of NASA-1.3~states~that~``\textit{Document any Modifications You make~to~the Licensed Software including the nature of the change, the authors of the change, and the date of the change}''. It belongs~to~$C_{g_6}$ as it requires to include \textit{Date}, \textit{Author} and \textit{Informative Statement}.
\end{exmp}

We analyze the distribution of \todo{47} open source licenses~across the \todo{six} groups of notice content. The result is shown in Fig.~\ref{fig:content}. \todo{13 (27.7\%)} licenses~belong~to~$C_{g_1}$ and $C_{g_3}$, respectively, and \todo{11 (23.4\%)} licenses~belong~to~$C_{g_6}$, which account for the largest portions~of~licenses. $C_{g_3}$ only requires to add~a~brief~statement,~which~is the simplest practice to obligate MTs among the \todo{six}~groups. Differently, $C_{g_6}$~is~the~strictest~practice~to~obligate MTs as~it requires date, author and informative statement. Each of the rest three groups contains no more than four licenses.


\begin{table*}[!t]
    \centering
    \scriptsize
    \setlength{\leftmargini}{0.4cm}
    \caption{\todo{RQ2:} Groups of Notice Content}\label{table:content_groups}
    \vspace{-10pt}
    \begin{tabular}{m{0.5cm}m{14cm}m{2.2cm}}
        \noalign{\hrule height 1pt}
        Group    & \MT~Excerpted from License Document & Composing Elements \\\hline
        $C_{g_1}$ &
        \textbf{LGPL-2.1}: You must cause the files modified to carry prominent notices \textbf{stating that you changed the files} and \textbf{the date of any change}.
                 & $C_{c_1}$, $C_{c_3}$                              \\
        \hline
        $C_{g_2}$ &
        \textbf{CDDL-1.0}: You must include a notice in each of Your Modifications that \textbf{identifies You as the Contributor of the Modification}.
                 & $C_{c_2}$, $C_{c_3}$                              \\
        \hline
        $C_{g_3}$ &
        \textbf{Apache-2.0}: You must cause any modified files to carry prominent notices \textbf{stating that You changed the files}.
                 & $C_{c_3}$                                        \\\hline
        $C_{g_4}$ &
        \textbf{LPPL-1.3c}: Every component of the Derived Work contains prominent notices \textbf{detailing the nature of the changes to that component}, or a prominent reference to another file that is distributed as part of the Derived Work and that contains \textbf{a complete and accurate log of the changes}.
                 & $C_{c_4}$                                        \\ \hline
        $C_{g_5}$ &
        \textbf{OGTSL}: You may otherwise modify your copy of this Package in any way, provided that you insert a prominent notice in each changed file \textbf{stating how} and \textbf{when you changed that file}...
                 & $C_{c_1}$, $C_{c_4}$                              \\\hline
        $C_{g_6}$ &
        \textbf{NASA-1.3}: Document any Modifications You make to the Licensed Software including \textbf{the nature of the change, the authors of the change}, and \textbf{the date of the change}.
                 & $C_{c_1}$, $C_{c_2}$, $C_{c_4}$                    \\
        \noalign{\hrule height 1pt}
    \end{tabular}
\end{table*}

\begin{table*}[!t]
    \centering
    \scriptsize
    \setlength{\leftmargini}{0.4cm}
    \caption{\todo{RQ3:} Groups of Notice Location}\label{table:location_groups}
    \vspace{-10pt}
    \begin{tabular}{m{0.5cm}m{14cm}m{2.2cm}}
        \noalign{\hrule height 1pt}
        Group    & \MT~Excerpted from License Document & Composing Elements \\\hline
        $L_{g_1}$ &
        \textbf{AFL-3.0}: You must cause all Covered Code to which You contribute to \textbf{contain a file documenting the changes} You made to create that Covered Code and the date of any change.
                 & $L_{c_1}$                                        \\\hline
        $L_{g_2}$ &
        \textbf{Apache-2.0}: You must \textbf{cause any modified files to carry prominent notices} stating that You changed the files.
                 & $L_{c_2}$                                        \\\hline
        $L_{g_3}$ &
        \textbf{RPL-1.1}: Document any Modifications You make to the Licensed Software including the nature of the change, the authors of the change, and the date of the change. This documentation must appear both \textbf{in the Source Code and in a text} file titled ``CHANGES'' distributed with the Licensed Software and Your Extensions.
                 & $L_{c_1}$,\ $L_{c_2}$                             \\\hline
        $L_{g_4}$ &
        \textbf{GPL-3.0}: The work must carry \textbf{prominent notices} stating that you modified it, and giving a relevant date.
                 & $L_{c_1}\ | \ L_{c_2}$                             \\\hline
        \noalign{\hrule height 1pt}
    \end{tabular}
\end{table*}

\textbf{Notice Location (\todo{RQ3}).} We summarize \todo{two} composing elements~of notice location (denoted as $L_{c_i}$, $1 \le i \le 2$), i.e.,~\textit{A~Document} and~\textit{Each Modified File}, as listed in Table~\ref{table:location_component}.~$L_{c_1}$ denotes~that notice content is added to a separate file.~$L_{c_2}$ denotes that~notice content is separately added to each modified file.


Based on these composing elements, we categorize licenses into \todo{four} groups (denoted as $L_{g_i}$, $1 \le i \le 4$) with respect~to the notice~location of their MTs, as shown in Table~\ref{table:location_groups}. Notice that~we~use~$L_{c_1}~|~L_{c_2}$ to denote that notice content can be added to either a document or each modified file if not clearly~stated.


\begin{exmp}\label{exmp:location_group}
The MT of AFL-3.0 states that ``\textit{You must~cause~all Covered Code to which You contribute to contain~a~file~documenting the changes You made to create that Covered~Code~and the date~of~any change}''. It belongs to $L_{g_1}$ as it requires~a~file to document the changes. The MT of RPL-1.1 states that~``\textit{Document any Modifications You make~.... This documentation must appear both in the Source Code and in a text file titled `CHANGES' ...}''. It belongs to $L_{g_3}$ as it requires to add notice content to both a file and each modified file. The MT of GPL-3.0 states that ``\textit{The work must carry prominent notices stating that you modified it, and giving a relevant date}''. It belongs to $L_{g_4}$ as it has no specific requirement for notice location.
\end{exmp}

We analyze the distribution of \todo{47} open source licenses~across the \todo{four} groups of notice location. The result is shown in Fig.~\ref{fig:location}.  $L_{g_2}$ and $L_{g_1}$ occupy the largest portion of licenses, accounting for \todo{42.6\%} (\todo{20}) and \todo{40.4\%} (\todo{19}) of licenses,~respectively.~This~indicates~that~it~is the most common requirement to put notice either in~a~separate~file or in each modified file.~\todo{7}~(\todo{14.9\%})~licenses (belonging to $L_{g_4}$) have no specific requirement for notice location, and only \todo{1} (\todo{2.1\%}) license (belonging to $L_{g_3}$) requires to put notice~in~both~a~separate~file~and~each~modified~file.


\textbf{Obligation Modeling (\todo{RQ4}).} For each of the \todo{47} open source~licenses, we model its MT $t$ as a 2-tuple $\langle lic, O \rangle$,~where~$lic$~denotes the license~name, and~$O$ denotes the obligation~of~$t$.~We model~$O$ as a 3-tuple~$\langle S_{g_i}, C_{g_j}, L_{g_k} \rangle$,~where~$S_{g_i}$,~$C_{g_j}$~and~$L_{g_k}$ are the modification scope, notice content and notice~location. 

Based~on~the~obligation, we categorize the \todo{47} licenses~into \todo{22} groups (denoted as $T_{g_n}$, $1 \le n \le 22$), where~each~group has~the~same obligation (i.e., $S_{g_i}$, $C_{g_j}$ and $L_{g_k}$). Fig.~\ref{fig:term_group}~shows the distribution of \todo{47} open source~licenses across these~groups. The \todo{nine} groups~containing the most licenses are separately reported, while the~\todo{13} groups containing only one license~are~put into \textit{Others}. $T_{g_1}$requires $S_{g_8}$,~$C_{g_1}$ and $L_{g_2}$, accounting for \todo{9 (19.1\%)} licenses. $T_{g_2}$ requires $S_{g_5}$, $C_{g_6}$~and~$L_{g_1}$, containing~\todo{6 (12.8\%)} licenses. Overall, nearly \todo{72.3\%} of the licenses declare the same obligation with at least one license.

\begin{figure*}[!t]
    \centering
    \includegraphics[width=0.9\linewidth]{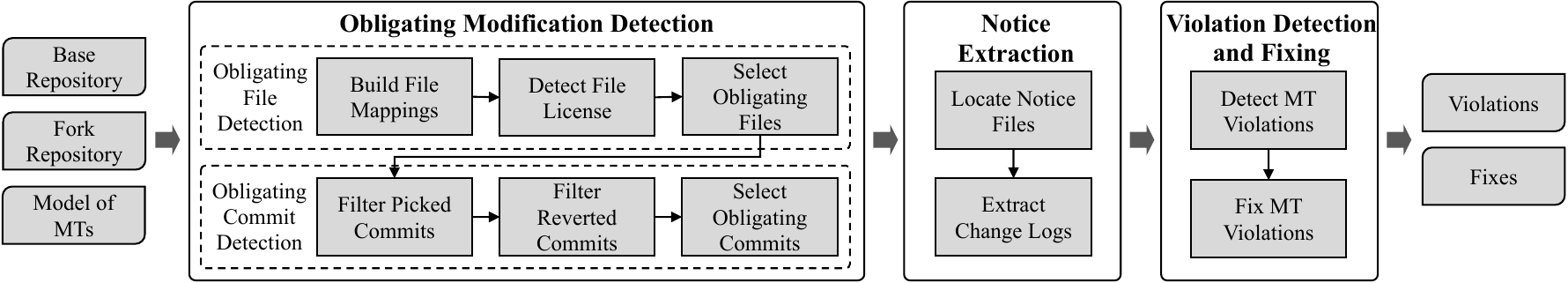}
    \vspace{-10pt}
    \caption{Approach Overview of \tool}\label{fig:overview}
\end{figure*}

\section{Approach}\label{sec:approach}

Inspired by our empirical study in Sec.~\ref{sec:understanding}, we propose~\tool to~automatically detect and~fix~violations of MTs during forking. We first introduce its overview, and then elaborate each step.


\subsection{Approach Overview}

Fig.~\ref{fig:overview} presents the approach overview of \tool. It takes as inputs~a base and a fork repository as well as the model~of~MTs for the \todo{47} open source licenses (Sec.~\ref{sec:understanding}).~\tool~works~in~three steps to detect~and fix violations of MTs (i.e., the required notice is missing~for~the~modified files that are in the modification scope) during forking. First, it detects obligating modifications of~the~fork~repository (Sec.~\ref{sec:approach:modification_detection}). Here we use commits~as the granularity to recognize~modifications. In that sense, we detect obligating commits that modify~the~original files that~are~in the modification scope of the MT~of~the~base~repository. Second,~it extracts change logs from potential notice location~in the~fork repository~(Sec.~\ref{sec:approach:changelog_localization_extration}).~These~change~logs~contain~potential~notice for the obligating commits. Finally,~it~matches change logs with commit logs based on notice content~of~the MT to detect and fix MT violations for the fork repository (Sec. \ref{sec:approach:report_generation}).

\subsection{Obligating Modification Detection}\label{sec:approach:modification_detection}


This step recognizes the fork repository's modifications on the base repository that are under the obligation~of~the~MT.~Here, we~choose commits as the granularity to detect such obligating modifications because commit logs provide sufficient semantic information about modifications which can ease the detection and fixing of MT violations. Therefore, this step aims to identify obligating commits~in~the fork repository. Each~obligating~commit modifies at least one obligating file (which is under the protection of the MT) in the base repository. To achieve this goal, this step is divided into two stages. The first is to detect obligating files protected by the MT. The second is to detect obligating commits which modify the obligating~files.

\subsubsection{Obligating File Detection}  After forking occurs,~the~base repository $B_1$ is evolved into the latest~base~repository~$B_2$. Meanwhile, the fork repository is evolved from $B_1$ into~$F$.~The modifications between $B_1$ and $F$ may change files in the~base repository which are protected by MTs. This stage builds~file mappings between the base and fork repository~to~denote~file reuse relations, detects the license of the file in the base~repository, and finally selects obligating files protected by the license.

\textbf{Build File Mappings.}  First, we build~file mappings between the base and fork repository.~Each~file~mapping is denoted~as $\langle f, f^{\prime} \rangle$, where $f \in B_1 \cup B_2$, $f^{\prime} \in F$, and~$f^\prime$ reuses~$f$.~Specifically, we build file mappings based on three reuse scenarios. First, for each file $f^\prime \in F$, if there exists $f \in$ $B_1 \cup B_2$ that has the same path name and file name with $f^\prime$~and is created by the base repository, we build~a~mapping~$\langle f, f^{\prime} \rangle$.~It indicates that the fork repository reuses $f$ and does not move~or rename $f$. Here we also consider files in $B_2$ because the fork repository may pick or merge commits from the base repository. Second, for each of the remaining files $f^\prime$~in~$F$,~we~backtrack the commit history, use the rename detector provided~by JGit~\cite{jgit} to locate the file $f$ in the base repository that~is~renamed~into~$f^\prime$, and build a mapping $\langle f, f^{\prime} \rangle$. It indicates that the fork repository reuses $f$ and moves~or renames $f$. Third,~for the rest files~in~$F$, we use \textsc{Saga}~\cite{li2020saga} to detect file-level clones against~$B_1$~and~$B_2$. If $f^{\prime} \in F$ has a clone relation with $f \in B_1 \cup B_2$ and~$f$~is~created by the base repository, we build a mapping $\langle f, f^{\prime} \rangle$. It indicates that the fork repository reuses the code in $f$ by copy-and-paste.

\textbf{Detect File License.} Then, for each file mapping $\langle f, f^{\prime} \rangle$,~we apply \textsc{Ninka}~\cite{german2010sentence} to detect the file license~of~$f$,~which~extracts text from file header and identifies the license~by~matching~extracted text against predefined license templates.~If~\textsc{Ninka}~reports an unknown result (i.e., no license is matched), we apply \textsc{Ninka} to identify project license, and use it as the file license. We use $lic_f$ to denote the detected file license of $f$.

\textbf{Select Obligating Files.} Last, for each file mapping $\langle f, f^{\prime} \rangle$, if the detected file license $lic_f$ belongs to the \todo{47} open source licenses, we determine whether $f$ is protected~by~the~MT~$t$~of~$lic_f$ ($t.lic = lic_f$), i.e., whether $f$ is in the modification scope of $t.lic$.~To~this~end, we distinguish $S_{c_1}$ to $S_{c_7}$ with file~extensions. Specifically,~we first search FileInfo.com (which is a database~of over 10,000~file extensions with detailed information about the associated file types) with names of $S_{c_1}$ to $S_{c_7}$~as~queries.~Then, we manually check and correct~the~query results. Finally,~we map \todo{2,085} file extensions to $S_{c_1}$ to $S_{c_7}$, and provide the detailed mappings at our website~\cite{zenodo}. Given such mappings,~if~the file extension of $f$ belongs to the corresponding file extensions of the modification scope~(i.e.,~$t.O.S_{g_i}$),~we~select~$f$~as~the~obligating file that is protected by $t$. We denote all obligating~files~as $F_{ob}$, and denote each obligating file as a tuple $\langle f, f^{\prime}, t \rangle$,~meaning that the obligating file $f$ is under the protection of $t$ and the modifications in $f^\prime$ made on $f$ should fulfill $t$.

\subsubsection{Obligating Commit Detection} We denote the commit~history of the base repository between $B_1$ and $B_2$~as~$H_B$,~denote the commit history of the fork repository between $B_1$ and $F$ as $H_F$, and denote each commit as a tuple $\langle id, a, d, msg \rangle$, where $id$ denotes the commit SHA id, $a$ denotes the author, $d$ denotes the date of the commit, and $msg$ denotes the commit log.~This stage first filters picked and reverted commits~(which~are free from the obligation) from $H_F$, and~then~selects~obligating~commits (which modify the obligating files in $F_{ob}$) from $H_F$.

\textbf{Filter Picked Commits.} The fork repository might~pick~or merge commits from the base repository. Such commits~are~submitted by the base repository, and hence~are~free~from~the~obligation. Therefore, we need to filter~such~commits~from~$H_F$.~In particular, for each commit $h \in H_F$, if there exists $h^\prime \in H_B$ that has the same SHA id with $h$ (i.e., $h.id = h^\prime.id$),~we~remove $h$ from $H_F$ (i.e., $H_F = H_F -\{h\}$). 

\textbf{Filter Reverted Commits.} The \texttt{git revert} command can be used to revert the changes introduced by a commit~$h$ and append a new commit $h^\prime$ with the resulting~reverse~content. As $h$ and $h^\prime$ do not produce any changes in the latest~fork~repository, we need to filter these reverted commits~from~$H_F$.~Specifically, the commit log of $h^\prime$ is automatically generated by the revert command with a pattern of ``Revert $h.msg$''. We use~this commit log pattern to identify the reverted commits~$h$ and $h^\prime$, and remove them from $H_F$ (i.e., $H_F = H_F -\{h, h^\prime\}$).

\textbf{Select Obligating Commits.} After filtering picked~and~reverted commits, we start with the latest commit $h \in H_F$~and~determine whether $h$ is an obligating commit by the following~procedure. We get the changed files in $h$, and~iterate~each~changed file. If a changed file $f_c$ is added,~deleted~or~modified and there exists an obligating file $f_{ob}$ $ \in F_{ob}$ such that~$f_c =$ $f_{ob}.f^\prime$~(i.e.,~$h$ modifies the obligating file), we regard $h$~as~an~obligating~commit, and add the required MT $f_{ob}.t$ to $T_h$. If a changed file is renamed from~$f_c$~to~$f_c^\prime$~and~there~exists an obligating file $f_{ob}$ $ \in F_{ob}$ such that~$f_c^\prime =$ $f_{ob}.f^\prime$ (i.e.,~$h$~renames the obligating~file), we regard $h$~as~an~obligating~commit, add the required MT $f_{ob}.t$ to $T_h$, and add $\langle f_{ob}.f, f_c, f_{ob}.t \rangle$ to $F_{ob}$ to continue~the~tracking of earlier changes on $f_c$. Then, we continue the above procedure on $h$'s previous commit in $H_F$.  Finally, we denote the set of identified obligating commits as $H_{ob}$, while the other commits between $B_1$ and $F$ are considered as obligation-free (\texttt{OF}).

For each obligating commit $h$, $T_h$ stores all the required MTs because $h$ might modify multiple obligating files protected by different MTs. Therefore, we conduct a union operation on the notice content and notice location required by the MTs~in~$T_h$,~as formulated by Eq.~\ref{eq:equation}, where $C_h$ and $L_h$  respectively denote~the required notice content and notice location for $h$.
\begin{equation}\label{eq:equation}
\begin{aligned}
            C_{h} &=  \cup_{t}^{T_{h}}(t.O.C_{g_j}),~L_{h} = \cup_{t}^{T_{h}}(t.O.L_{g_k})
\end{aligned}
\end{equation}
For example, given an obligating commit that changes~two~obligating files $f_1$ and $f_2$, $f_1$ conforms to $t_1$, and $f_2$ conforms~to~$t_2$. $t_1$ requires $C_{c_3}$ as the notice content, and $t_2$ requires $C_{c_1}$ and $C_{c_4}$ as the notice content. Thus, the required notice content~for this obligating commit is $C_{c_1}$, $C_{c_2}$ and $C_{c_4}$.

\subsection{Notice Extraction}\label{sec:approach:changelog_localization_extration}

This step locates notice files in $F$,~and~extracts change logs which could be notice for obligating~commits.

\textbf{Locate Notice Files.} We locate notice files in $F$ based on our understanding of notice location (see Sec. \ref{sec:mt-modeling}) which can be either a separate document (i.e., $L_{c_1}$) or each modified~file~(i.e., $L_{c_2}$). On the one hand, regarding $L_{c_1}$, we search~files~in~$F$~with certain file name patterns, e.g., release.txt and ChangeLog.txt. We provide the full list of file name patterns at our website~\cite{zenodo}. On the other hand, we do not analyze files~regarding~$L_{c_2}$~since \todo{developers often do not put notice in each modified file, which is evidenced when we prepare our ground truth (see Sec.~\ref{sec:setup})}. Thus, we relax the obligation of adding notice to each modified file to allow for notice in a separate document.


\begin{figure}[!t]
    \centering
    \fbox{\includegraphics[width=0.8\linewidth]{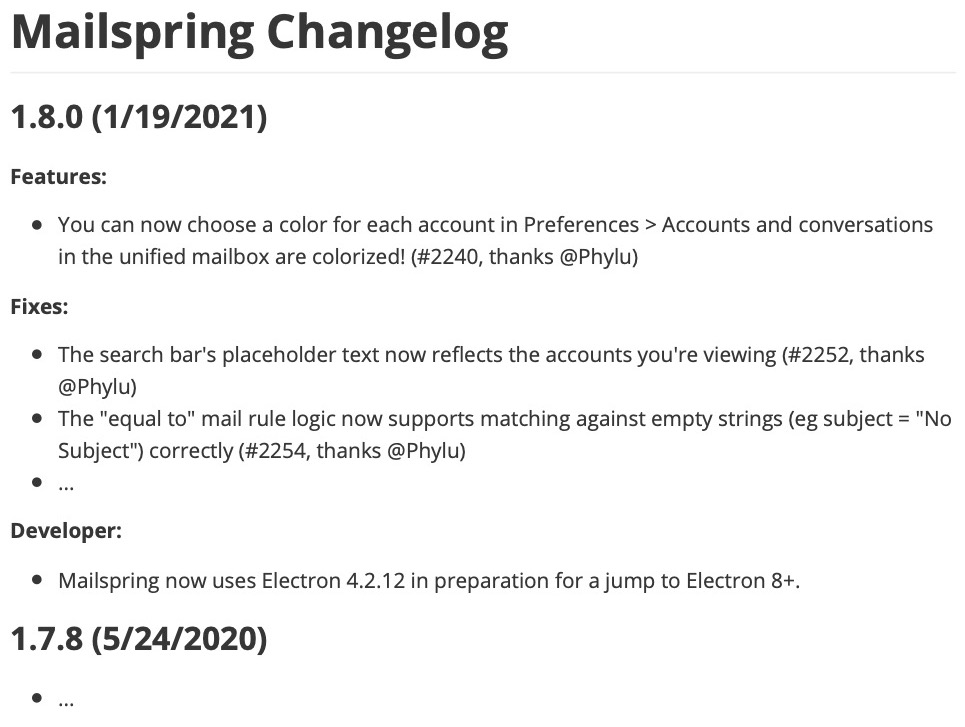}}
    \vspace{-10pt}
    \caption{An Illustrative Example of Notice Files}\label{fig:date_demo}
\end{figure}

\textbf{Extract Change Logs.} After we locate notice files,~we~parse and extract change logs. Generally, change logs are written in semi-structured texts whose extraction can be automated based on templates. To derive the templates,~we~manually inspect~\todo{519} fork repositories from the \todo{1,747} fork repositories that have~over 300 stars and have development activities~(e.g.,~commits)~after October, 2018, achieving a confidence level~of~\todo{99\%} and~a~margin of error of \todo{4.76\%}. We successfully locate notice files~in~\todo{100} fork repositories. Specifically, most~of~them~(i.e.,~\todo{91}~fork~repositories) have change logs organized in blocks. Each block~consists of a subtitle and a list of change logs. The subtitle contains date literals in \todo{55} fork repositories, and has only version literals in the rest \todo{36} fork repositories. Fig.~\ref{fig:date_demo} shows a clip of notice file in Mailspring-Libre\footnote{\scriptsize{https://github.com/notpushkin/Mailspring-Libre/blob/master/CHANGELOG.md}}, where the subtitle contains both date and version literals. Besides, the rest \todo{9} fork repositories have short change logs which are not separated~by~subtitles.

\begin{figure}[!t]
    \centering
    \includegraphics[width=0.8\linewidth]{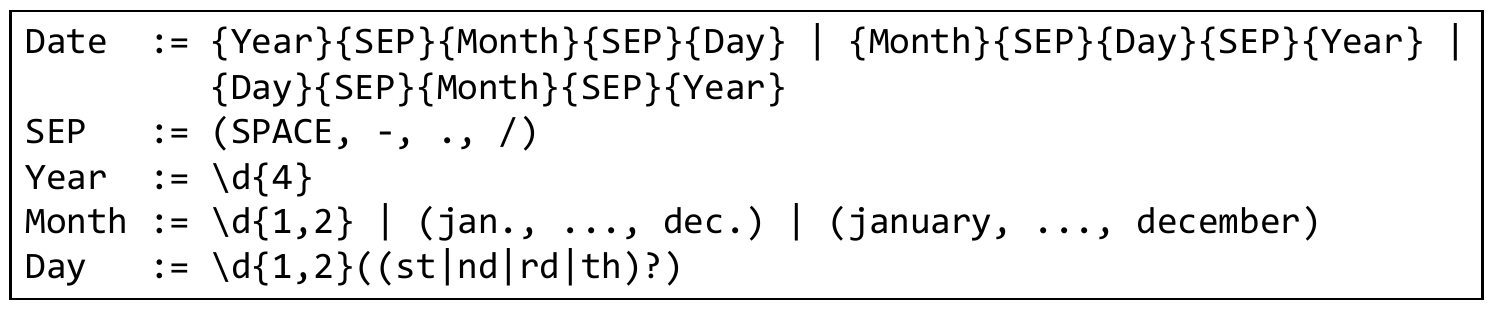}
    \vspace{-10pt}
    \caption{The Pattern of Date}\label{fig:date_pattern}
\end{figure}

For notice files where change logs are separated by subtitles, we iterate each line for date expressions, which~are~matched~by the pattern in Fig.~\ref{fig:date_pattern}. Each date expression is the starting~point~of a block of change logs. Each block is associated with~a~range~of date, i.e., from the starting date~$d_s$~to~the~ending~date~$d_e$,~indicating that the modifications described by the change logs~in~the block occur between $d_s$ and $d_e$. Here, $d_e$ is extracted from the current block, while $d_s$ is extracted from~the~next block. For example, $d_s$ and $d_e$ of the first block in Fig.~\ref{fig:date_demo} are respectively 2020-05-24 and 2021-01-19. Then, we consider each non-blank line $msg$ in the block as a change log. For each non-blank~line, we search for author $a \in A$ ($A$~is~obtained~by~collecting author information from each commit in $H^F$ before filtering).~For~example, Phylu is the author for three change~logs~in~Fig.~\ref{fig:date_demo}. For other notice files (i.e., with subtitles that do not contain data expressions, and with no subtitles), we consider each non-blank line as a change log in the same way as the above procedure.

We denote a change log $n$ as a tuple $\langle msg, d_s, d_e, a\rangle$,~and~denote the whole set of change logs as $N$. $d_s$, $d_e$~and $a$ might be blank when there is no date expression and author declaration~in change logs. Note that the extracted change logs might be noisy. For example, the non-blank line ``Features:''~in~Fig.~\ref{fig:date_demo}~is~also~considered~as a change log. Besides, some change logs describe~the changes of fork repositories' own files. However, such noises will not affect the accuracy of \tool because such noisy change logs will not be matched to any obligating commits.

\subsection{Violation Detection and Fixing}\label{sec:approach:report_generation}

Given obligating commits $H_{ob}$ and change logs $N$, this step detects MT violations and proposes fixes for MT violations.

\textbf{Detect MT Violations.} For each obligating commit $h \in H_{ob}$, we try to find a change log $n \in N$ such that $n$ fulfills~the~obligation required by the modification in $h$. If such a change~log~is not found, we detect a MT violation for $h$.

Specifically, if the notice content $C_{c_3}$ or $C_{c_4}$ is required,~i.e., $C_{c_3} \in C_{h}$ or $C_{c_4} \in C_{h}$ (in fact every MT of the~\todo{47}~open~source licenses requires $C_{c_3}$ or $C_{c_4}$), we measure~the~similarity~between the commit~log $h.msg$ and the change~log~$n.msg$.~If~the similarity exceeds a threshold $th$, the obligating commit~$h$~is regarded to fulfill $C_{c_3}$ or $C_{c_4}$ through the change log $n$.~If~such a change log is not found, we detect a MT violation~for~$h$~due to missing notice (\texttt{VN}). Here, we do not distinguish between $C_{c_3}$ (\textit{brief statement}) and $C_{c_4}$ (\textit{informative statement}) as~the~boundary between brief and informative~natural~language~texts~lacks a clear definition. We use TF-IDF~\cite{luhn1957statistical, jones1972statistical}~to~measure~the~similarity between $h.msg$ and $n.msg$. $h.msg$ and $n.msg$~are~represented by a term frequency vector, and the document collection is the list of change logs in $N$. We implement the TF-IDF by using \textit{TfidfTransformer} from \textit{scikit-learn}~\cite{scikit-learn}. 


Then, if such a change log $n$ is found and the notice~content $C_{c_1}$ is required (i.e., $C_{c_1} \in C_{h}$), we compare the date~of~$h$~with the date range of $n$. If $h.d$ is within the date range~of~$n.d_s$~and $n.d_e$, the obligating commit $h$ is considered to fulfill~$C_{c_1}$.~Otherwise, we detect a MT violation due to missing date (\texttt{VD}). 

Finally, if such a change log $n$ is found and the notice~content $C_{c_2}$ is required (i.e., $C_{c_2} \in C_{h}$), we compare~the~author~of~$h$ with the author of $n$. If $h.a$ is the same as $n.a$, the obligating commit $h$ is considered to fulfill $C_{c_2}$. Otherwise, we detect a MT violation due to missing author (\texttt{VA}). If $h$ fulfills all the obligation in $C_{h}$, it is considered as fully obligated (\texttt{OB}).

Our detection result for each obligating commit $h \in H_{ob}$~is denoted as a tuple $\langle h, n, type\rangle$, where $h$ denotes the obligating commit, $n$ denotes the matched change log,~and~$type$~denotes the fulfillment type which can be either violated due~to~missing notice (\texttt{VN}), violated due to missing date (\texttt{VD}), violated~due~to missing~author (\texttt{VA}), and fully obligated (\texttt{OB}).

\textbf{Fix MT Violations.} For an obligating commit $h$ whose~fulfillment type is \texttt{VN}, we create a new change log $n$ such that $n.msg=h.msg$, $n.d_s=h.d$, $n.d_e=h.d$ and $n.a=h.a$. For an obligating commit $h$  whose fulfillment type is \texttt{VD} (resp.~\texttt{VA}), we update the matched change log $n$ such that $n.d_s=h.d$ and $n.d_e=h.d$ (resp. $n.a=h.a$). Finally, we write the newly-added or updated change log to the notice file. If the notice~file does not exist, we also create a new notice file.

\section{Evaluation}\label{sec:experiments}

We have implemented \tool with \todo{4.2K} lines of \todo{Java} code~and \todo{0.4K} lines of Python code, and released the source code~at our website \cite{zenodo} with all of our experimental data.

\subsection{Evaluation Setup}\label{sec:setup}

To characterize real-world practices of MTs and evaluate~the~effectiveness and efficiency of \tool,~we~design~three~research~questions.

\begin{itemize}[leftmargin=*]
    \item \textbf{\todo{RQ5} MT Prevalence Evaluation}: How is the prevalence~of MTs in real-world fork repositories? (see Sec.~\ref{sec:experiments:rq1})
    \item \textbf{\todo{RQ6} MT Violation Evaluation}: How is the violation~of~MTs in real-world fork repositories? (see Sec.~\ref{sec:experiments:rq2})
    \item \textbf{\todo{RQ7} Effectiveness and Efficiency Evaluation}: How is the effectiveness and efficiency of \tool? (see Sec.~\ref{sec:experiments:rq3})
\end{itemize}


\textbf{Repository Collection.} We use Github GraphQL Explorer to query and collect pairs of base and fork repositories.~Specifically, we query for fork repositories that have over 300 stars~and have development activities (e.g., commits) after October, 2018. Our query returns \todo{1,743} fork repositories~with~their~corresponding base repositories. \todonew{Then, we identify \todo{814} fork repositories whose licenses are within the list of the \todo{107} licenses in our study in Sec.~\ref{sec:understanding}. Of the rest \todo{929} repositories, \todo{884} repositories do not declare any license, and \todo{45}~repositories declare \textit{unlicense}~\cite{unlicense} or other licenses which are out of the scope of our \todo{107} licenses.} \todonew{Further, we filter \todo{117} fork repositories from the \todo{814} fork repositories because \todo{84} of them indicate through ReadMes or merged commits that they have collaboration relations with the base repositories, \todo{12} of them are no longer maintained, and \todo{21} of them do not disclose the organization information so that the relation of developers between the base and fork developers is unknown.} \todonew{This results in \todo{697} fork repositories. Finally, we obtain \todo{178} fork repositories whose licenses are within the scope of our \todo{47} open source licenses while the rest are not.}



We further manually locate and analyze notice files in these \todo{178} fork repositories. We find that \todo{57} (\todo{32.0\%}) of these~\todo{178}~fork repositories contain notice files, which is relatively small. These \todo{57} fork repositories have an overlap of \todo{18} with the sampled \todo{519} fork repositories in Sec.~\ref{sec:approach:changelog_localization_extration}. All of these \todo{57} fork repositories have change logs organized in blocks. Among them, \todo{47} fork repositories have subtitles inclusive of date expressions, and~\todo{10} fork repositories have subtitles absent of date expressions.


\begin{figure}[t]
    \centering
    \includegraphics[width=0.72\linewidth]{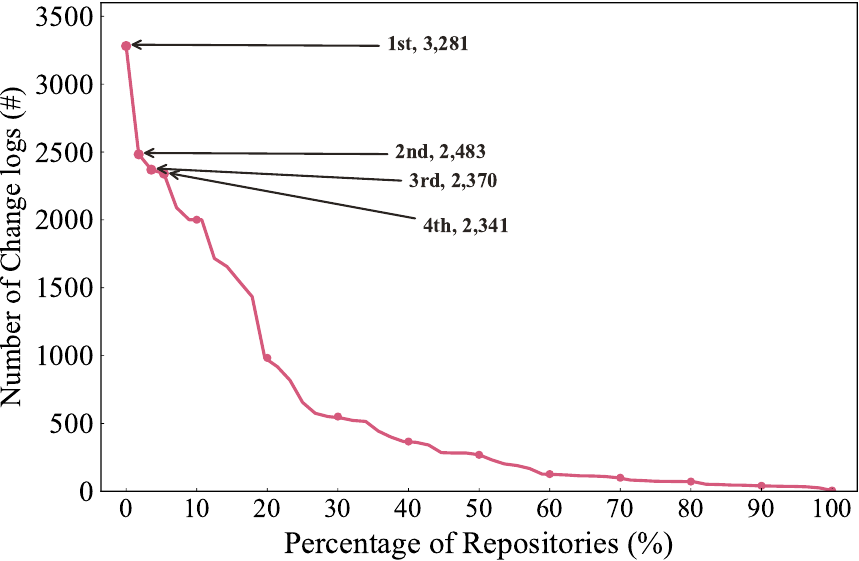}
    \vspace{-10pt}
    \caption{Number of Change Logs in Fork Repositories}\label{fig:rq1_changelog}
\end{figure}

We also manually extract change logs in these \todo{57} fork~repositories, rank the fork repositories in the descending order~of~the number of change logs, and plot them in Fig.~\ref{fig:rq1_changelog}. \todo{70\%}~of~the \todo{57} fork repositories have less than \todo{551} change logs. The maximum number of change~logs~in~a~single fork repository is \todo{3,281}, but this number decreases drastically in the following fork repositories, with \todo{2,483} change logs in the second,~\todo{2,370}~change logs in the third, and \todo{2,341} change logs in the fourth fork repository. These four fork repositories contribute \todo{30.4\%} of all the \todo{34,487} change logs in these \todo{57} fork repositories. 


\textbf{RQ Setup.} To explore \textbf{\todo{RQ5}}, we analyze the distribution~of modification scope, notice content, notice location and obligation group across the \todo{178} base repositories according to the licenses detected by \textsc{Ninka}. To study \textbf{\todo{RQ6}},~we~run~\tool~against each pair of base~and fork repositories, and obtain~the~detection result of MT violations. Specifically, we measure~the~number~of repositories~and commits that violate MTs. 
To investigate~\textbf{\todo{RQ7}}, we evaluate the accuracy, usefulness and performance~of~\tool. For accuracy, we manually construct the ground truth, and measure the detection accuracy of \tool on~different~fulfillment types. Since it is a multi-class classification problem,~we use macro-precision and macro-recall~\cite{grandini2020metrics} as the~accuracy~metrics. Further, we measure the accuracy sensitivity of \tool to the configurable similarity threshold $th$. For usefulness,~we~submit pull requests with detailed reference on violations and fixes to the corresponding fork repositories.~For~performance,~we~measure the~time~overhead~of~\tool on each repository pair.

\textbf{Ground Truth Construction.} We randomly sample~and~manually check commits from a total of \todo{176,273} commits in the~\todo{178} fork repositories until we successfully identify \todo{100} commits~for each fulfillment type (i.e., \texttt{OB}, \texttt{VN} and \texttt{VD})~as~well~as~\todo{100}~commits that are obligation-free (i.e., \texttt{OF}). We fail to identify any commit for \texttt{VA} because no repository requires $C_{c_2}$ (i.e., author) in their licenses. Finally, we sample a total of \todo{39,022} commits, and construct a ground truth for~\todo{400}~commits,~achieving~a~confidence level of \todo{95\%} and a margin error of \todo{4.9\%}. Specifically, two of the authors independently determine the fulfillment type for each sampled commit. We use Cohen’s Kappa coefficient~to measure agreement, and it reaches \todo{0.891}. A third author is involved to resolve disagreements. Notice that the ground truth is manually constructed in two person-months.

\begin{figure*}[!t]
    \centering
    \begin{subfigure}[b]{0.24\textwidth}
        \centering
        \includegraphics[width=0.99\textwidth]{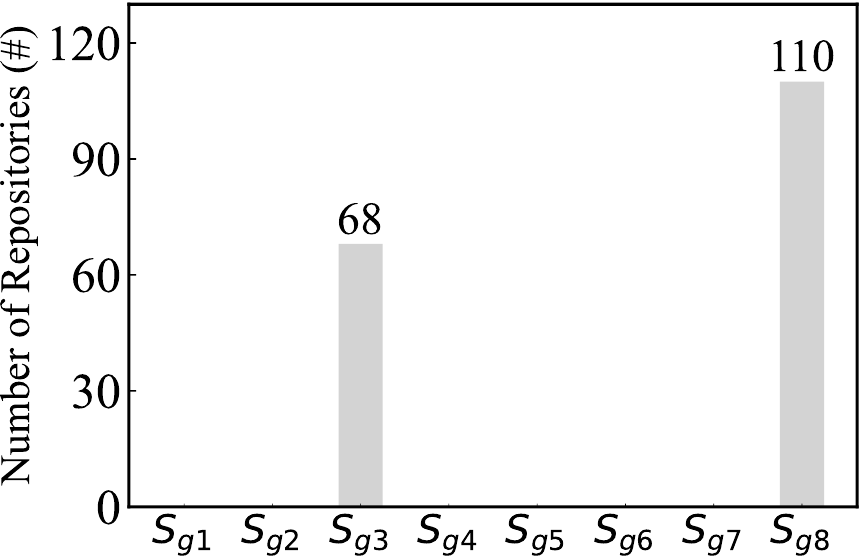}
        \vspace{-10pt}
        \caption{Modification Scope}\label{fig:rq_scope}
    \end{subfigure}
    \begin{subfigure}[b]{0.24\textwidth}
        \centering
        \includegraphics[width=0.99\textwidth]{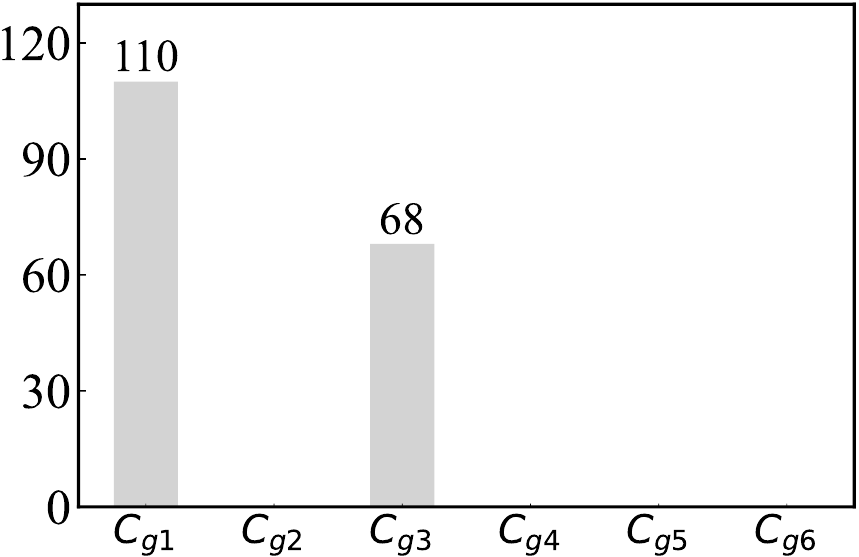}
        \vspace{-10pt}
        \caption{Notice Content}
        \label{fig:rq_content}
    \end{subfigure}
    \begin{subfigure}[b]{0.24\textwidth}
        \centering
        \includegraphics[width=0.99\textwidth]{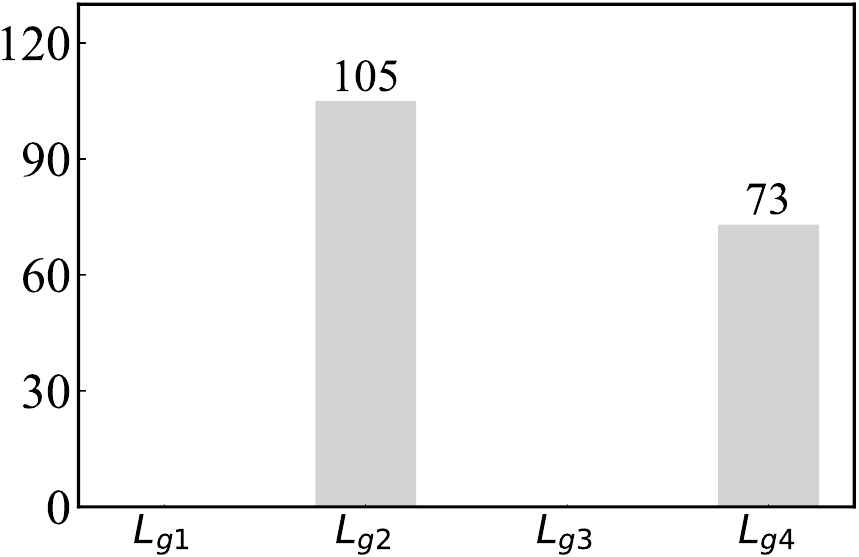}
        \vspace{-10pt}
        \caption{Notice Location}\label{fig:rq_location}
    \end{subfigure}
    \begin{subfigure}[b]{0.24\textwidth}
        \centering
        \includegraphics[width=0.99\textwidth]{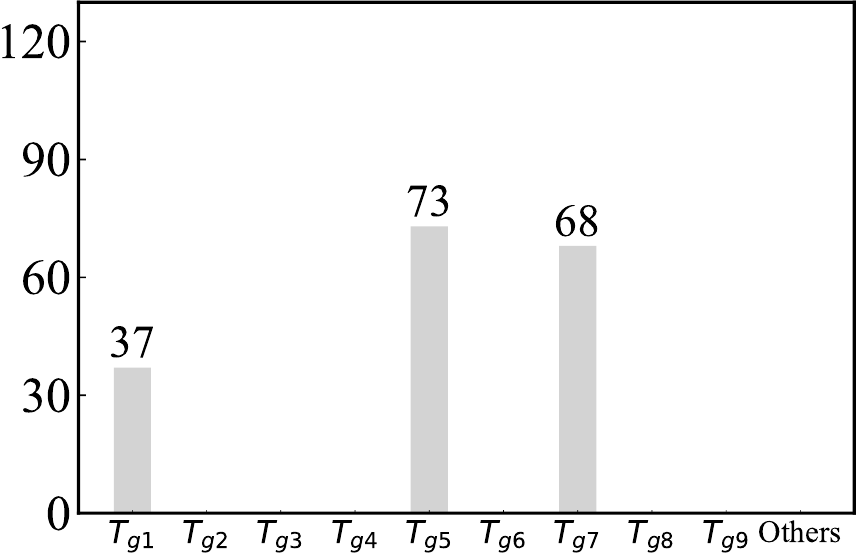}
        \vspace{-10pt}
        \caption{Obligation Group}\label{fig:group_all}
    \end{subfigure}
    \vspace{-10pt}
    \caption{Number of Repositories w.r.t Modification Scope, Notice Content, Notice Location and Obligation Group}\label{fig:obligation}
\end{figure*}

\begin{figure*}[t]
    \centering
    \begin{subfigure}[b]{0.32\textwidth}
        \centering
        \includegraphics[width=0.89\textwidth]{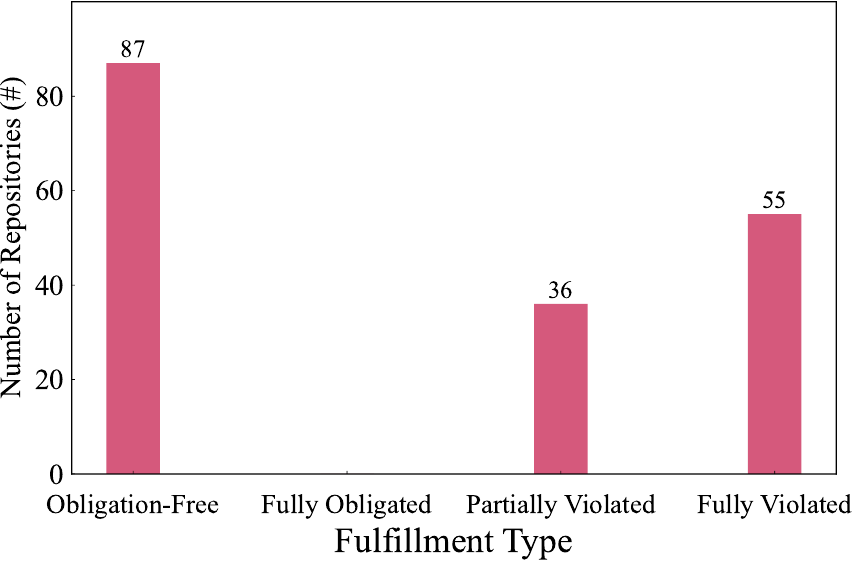}
        \vspace{-5pt}
        \caption{\todo{Repositories That Violate MTs}}\label{fig:non_obligating_commits}
    \end{subfigure}
    \begin{subfigure}[b]{0.32\textwidth}
        \centering
        \includegraphics[width=0.89\textwidth]{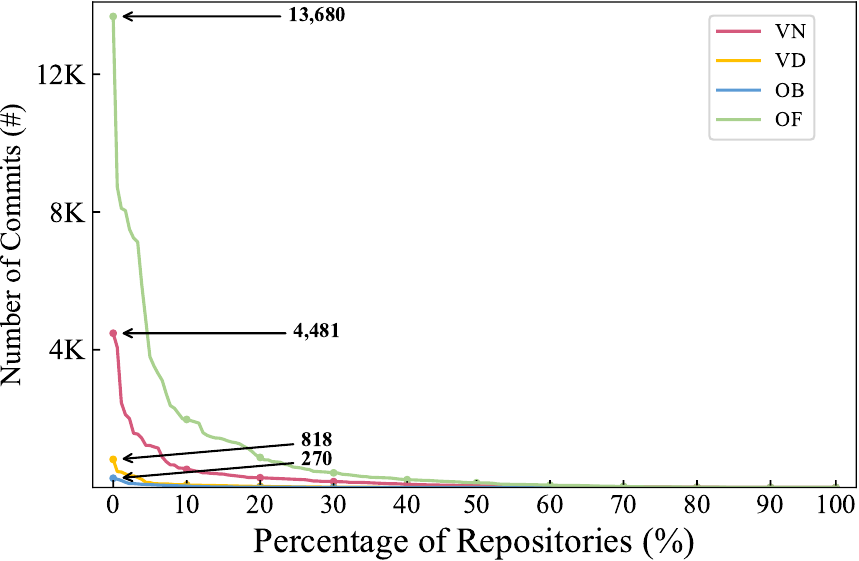}
        \vspace{-5pt}
        \caption{\todo{Commits That Violate MTs}}\label{fig:obligating_commits}
    \end{subfigure}
   \begin{subfigure}[b]{0.34\textwidth}
       \centering
       \includegraphics[width=0.89\textwidth]{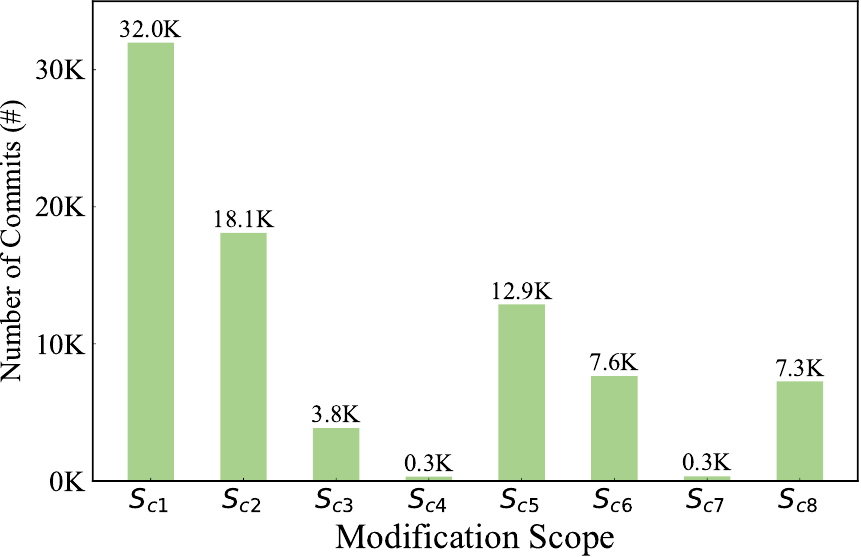}
       \vspace{-5pt}
       \caption{\todo{Obligating Commits w.r.t Modification Scope}}\label{fig:rq1_commit_scope}
   \end{subfigure}
    \vspace{-10pt}
    \caption{The Number of Repositories and Commits That Violate MTs}\label{fig:violation}
\end{figure*}

\subsection{Real-World MT Prevalence (\todo{RQ5})}\label{sec:experiments:rq1}




As revealed by our repository collection in Sec.~\ref{sec:setup}, \todo{178}~of the \todo{814} highly-starred forked repositories require MTs~in~their licenses, which stands for a non-negligible proportion.~In~that sense, it is useful to detect and fix MT~violations during~forking. Surprisingly,  these~\todo{178}~base~repositories~only~adopt five~dominating licenses, i.e., Apache-2.0, GPL-3.0, GPL-2.0, AGPL-3.0 and LGPL-2.1. These five licenses are respectively adopted in \todo{68}, \todo{59}, \todo{32}, \todo{14} and \todo{5} base repositories. Regarding MTs~in~these licenses, we report the number of repositories with~respect~to~the three obligation dimensions (i.e., modification scope, notice~content, and notice location) and the obligation groups in Fig.~\ref{fig:obligation}. 

In particular, Apache-2.0 has the obligation $\langle S_{g_3}, C_{g_3}, L_{g_2} \rangle$, therefore contributing the bar with \todo{68} repositories in Fig.~\ref{fig:rq_scope}~and Fig.~\ref{fig:rq_content}. The GPL family (i.e., GPL-3.0, GPL-2.0, AGPL-3.0 and LGPL-2.1) has the obligation of $S_{g_8}$ and $C_{g_1}$,~therefore~contributing the bar with \todo{110} repositories in Fig.~\ref{fig:rq_scope} and Fig.~\ref{fig:rq_content}. Meanwhile, the GPL family is divided into two sub-groups,~i.e., GPL-2.0 and LGPL-2.1 whose notice location is $L_{g_2}$, and GPL-3.0 and AGPL-3.0 whose notice location is $L_{g_4}$, respectively accounting for \todo{37} and \todo{73} repositories.~As~a~result, the MTs fall into three~obligation groups,~as~shown in Fig.~\ref{fig:group_all}. $T_{g_5}$~represents GPL-3.0~and AGPL-3.0, accounting for \todo{73} repositories. $T_{g_7}$ represents Apache-2.0, taking~up \todo{68} repositories. $T_{g_1}$ represents GPL-2.0 and~LGPL-2.1, accounting for \todo{37} repositories.

\textit{\textbf{Summary.}} \todo{21.9\%} of the highly-starred forked repositories~require MTs~in~their licenses, indicating the moderate prevalence of MTs during forking. Due to the concentrated~adoption~of~five licenses, the MT obligation is concentrated in three groups.


\subsection{Real-World MT Violations (RQ6)}\label{sec:experiments:rq2}



\textbf{Obligated/Violated Repositories.} We categorize fork repositories into \textit{Obligation-Free}, \textit{Fully Obligated}, \textit{Partially Violated} and \textit{Fully Violated} repositories. The result is reported~in~Fig.~\ref{fig:non_obligating_commits}. \textit{Obligation-Free} repositories mean that~they~have no obligating commit because they do not~modify the original~files~protected by MTs, taking up \todo{87} of the \todo{178} fork repositories. \textit{Fully~Obligated} repositories have obligating commits~and~all~of~them~fulfill the MTs by corresponding change logs.~Unfortunately, we find zero fork repository that falls into this category.~\textit{Partially Violated} repositories have obligating commits and part of them violate the MTs, taking up \todo{36} of the \todo{178} fork repositories. \textit{Fully Violated} repositories have all of their obligating commits violate the MTs, taking up \todo{55} of the \todo{178} fork repositories. Overall, \todo{91 (51.1\%)} fork repositories violate the \MTs.

\textbf{Obligated/Violated Commits.} \tool detects \todo{53,982} obligating commits from the \todo{178} fork repositories, which~is~significantly larger than the total number of \todo{34,487} change~logs~(as reported in Sec.~\ref{sec:setup}). Finally, \tool detects a total of \todo{51,435} MT violations, i.e., \todo{51,435} commits that violate the MTs. Only a small part (\todo{4.7\%}) of obligating commits fulfill required MTs.

We rank the fork repositories in the descending order~of~the number of commits that are obligation-free (\texttt{OF}), fully obligated (\texttt{OB}), violated due to missing notice (\texttt{VN}), and violated due to missing date (\texttt{VD}), and plot them in Fig.~\ref{fig:obligating_commits}. \tool does not detect any commit that is violated~due~to~missing~author (\texttt{VA}) as~no~repository~requires $C_{c_2}$ (i.e., author) in their~licenses.~We can see that the number of obligation-free commits~(\texttt{OF})~significantly exceeds the number of obligating commits (i.e., \texttt{OB}, \texttt{VN} and \texttt{VD}). This is reasonable because fork repositories often extend base repositories by adding new features. Among the obligating commits, there are respectively \todo{45,470} and \todo{5,965} violated commits due to missing notice (\texttt{VN}) and missing date (\texttt{VD}), which are \todo{18} and \todo{2} times larger than the \todo{2,547} obligated commits (\texttt{OB}). For example, the project \textit{INTI-CMNB/KiBot} has the largest number of obligated commits (\texttt{OB}), which is \todo{270}. Contrarily, the number of violated commits due to missing notice (\texttt{VN}) and missing date (\texttt{VD}) reaches their peak at \todo{4,481} in the project \textit{SonixQMK/qmk\_firmware} and \todo{818} in the project \textit{jobobby04/TachiyomiSY}, respectively.

\todonew{\textbf{Obligating Commits w.r.t Modification Scope.} Fig. \ref{fig:rq1_commit_scope} presents the number of obligating commits with respect to modification~scope. Specifically, source code ($S_{c_1}$) is modified in \todo{32.0K} commits, accounting for \todo{59.4\%} of the obligating commits, followed by documentation ($S_{c_2}$) with \todo{18.1K} commits, and scripts ($S_{c_5}$) with \todo{12.9K} commits.}


\textit{\textbf{Summary}}. MT violations are quite severe in real-world fork repositories. \todo{51.1\%} of the fork repositories violate MTs. \todo{95.3\%}~of~the obligating commits violate MTs. 
\todonew{\todo{59.4\%} of the obligating commits contain modification to source code ($S_{c_1}$), making it the most prevalent element compared to other modification scope elements.}



\subsection{Effectiveness and Efficiency of \tool (RQ7)}\label{sec:experiments:rq3}



\textbf{Accuracy.} Using~our~constructed ground truth, we report~the accuracy of \tool in detecting the four types of commits~(i.e., \texttt{VN}, \texttt{VD}, \texttt{OF} and \texttt{OB}) in Table~\ref{table:accuracy}.~We use the multi-class measure of precision and recall provided~in \textit{scikit-learn}~\cite{scikit-learn} to compute the result. Overall, \tool achieves a macro-precision of \todo{0.82} and a macro-recall of \todo{0.80}. Moreover,~we~analyze~the~false~positives and false negatives to summarize their reasons. The major reason is the semantic unawareness of TF-IDF~in matching commit logs and change logs. It can incur both false positives (e.g., falsely detecting semantic-nonequivalent logs as matched, especially when the logs are short) and false negatives (e.g., falsely detecting semantic-equivalent logs as unmatched). 

\begin{figure}[!t]
    \centering
    \includegraphics[width=0.69\linewidth]{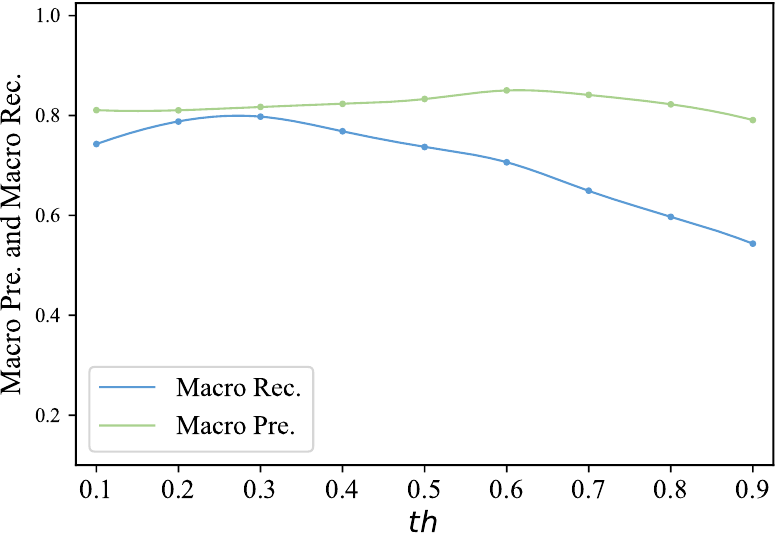}
    \vspace{-10pt}
    \caption{Accuracy of \tool w.r.t Similarity Threshold $th$}\label{fig:accuracy}
\end{figure}

Further, we analyze the sensitivity of the accuracy of \tool to the similarity threshold $th$. Specifically, we configure~$th$~from 0.1 to 0.9 by a step of 0.1, and report the result in Fig.~\ref{fig:accuracy}.~\tool achieves the best macro-precision and macro-recall when~$th$~is set to \todo{0.3}. As $th$ increases from \todo{0.3}, macro-precision~is~relatively stable, whereas macro-recall suffers a decreasing trend. Therefore, \tool is sensitive~to~$th$,~and~we~believe~\todo{0.3}~is~empirically a good value for $th$. Notice that we report the result in Table~\ref{table:accuracy} when $th$ is set to \todo{0.3}.



\begin{table}[!t]
    \centering
    \small
    \caption{Accuracy Results}\label{table:accuracy}
    \vspace{-10pt}
    \begin{tabular}{m{0.8cm}m{0.8cm}m{0.8cm}m{0.8cm}m{0.8cm}m{0.8cm}}
        \hline
        Metric & \texttt{VN} &   \texttt{VD} & \texttt{OF} & \texttt{OB} & Macro-Average \\\hline
        Pre. &  0.67  &  0.69   &  1.00    &  0.92  &  0.82    \\
        Rec. &  0.76  &  0.78   &  0.93    & 0.72   &  0.80    \\
        \hline
    \end{tabular}
\end{table}

	
\textbf{Usefulness.} To evaluate the practical usefulness of \tool,~we submit \todo{91} pull requests, using the violations and fixes generated by \tool, to the violated repositories which~have active~development activities recently in order to obtain~quick feedback.~So~far,~we have received responses from \todo{44} fork repositories.

In particular,  \todo{18} fork repositories give positive responses in the corresponding pull requests. \todo{8} fork repositories approve our pull requests and merge them into the main branch~of~their repositories. \todo{2} fork repositories fix the violations themselves using a separate commit, while \todo{1} fork repository is immediately archived to be read-only. \todo{4} fork repositories claim that they~are authors from the base repositories and hence there is no need to conform to such obligations. \todo{3} fork repositories~respond~positively but do not take further actions so far.

Despite the positive responses, we also receive unwillingness from \todo{26} fork repositories. \todo{7} of them directly close~the~pull~requests without further comments, while \todo{4} of them discuss the pull requests but do not reveal their attitude or actions.~\todo{15}~of them refuse to do further actions. Specifically, \todo{5} of them think that the commit history itself reflects the fulfillment of MTs and thus there is no violation. Interestingly, although \todo{2} of them refuse, they choose to delete or archive their repositories potentially for avoiding legal risks. \todonew{\todo{4} of them disagree with our~violation~definition. Particularly, the first one simply replies with a ``No'' without any further comment. The second one considers the statement of forking the base repository in the ReadMe file as fulfilling the MTs. However, the statement fails to describe the required notice content. The third one claims that only the original authors of the base repository can request to obligate the MTs. The last one disagrees with the violated commits.} Besides, \todo{4} of them consider our pull requests as spams.

\textbf{Performance.} To evaluate the performance of \tool,~we~measure the time cost for each pair of repositories. \tool takes~\todo{72.2} seconds in median and \todo{229.8} seconds on average for each~repository pair. For half of the repository pairs, \tool consumes~\todo{32.4} seconds to \todo{190.7} seconds. The minimal time cost is \todo{8.9} seconds, while the maximal time cost is \todo{4481.8} seconds. \todonew{Notice that our obligating modification detection step accounts for approximately \todo{92.3\%} of the overall time cost, primarily attributed to intensive Git operations.}

\textit{\textbf{Summary}}. \tool achieves a macro-precision of \todo{0.82}~and~a macro-recall of \todo{0.80}. \todo{18} pull requests of fixing modification term violations have received positive responses, and \todo{8} of them have been merged. These results demonstrate the effectiveness of \tool. Besides, \tool takes \todo{229.8} seconds on average~to~analyze one repository pair, which demonstrates its efficiency.

\subsection{Discussion}

\textbf{Threats.} 
The main threats to our evaluation is the ground~truth construction. First, the scale of the ground truth is not large~due to the expensive manual efforts. However, the scale ensures~an acceptable confidence level. Second, the quality~of~the~ground truth might affect our effectiveness evaluation. We mitigate~it~by following open coding procedure to build the ground truth.

\textbf{Limitations.} The limitations of our work are four-folds.~First, our model of MT obligations is not exhaustive to all licenses. We plan to include more licenses into our model. Second,~\tool currently takes base and fork repositories with their commit history as inputs. \todo{The violation may be more prevalent and more severe when commit history is not publicly available}. We plan to extend \tool to use software composition analysis (SCA) to detect obligating files, and code change summarization to generate logs. Third,~as~revealed~by~our~accuracy evaluation, the main inaccuracy is \todo{caused by semantic unawareness between $h.msg$ and $n.msg$ due to the limitations of TF-IDF.} We plan to leverage semantic-aware techniques like deep learning and use large-scale training data to further improve the accuracy of \tool. Finally, we currently employ some relaxation to not distinguish between $C_{c_3}$ and $C_{c_4}$ and between $L_{c_1}$ and $L_{c_2}$.~We plan to resort to license regulators for formal interpretations.




\section{Related Work}



\textbf{License Identification.} To facilitate the analysis of licenses, the first step is to automatically identify licenses from text~or code files. Tuunanen et al.~\cite{tuunanen2009automated} develop \textsc{ASLA} to identify licenses based~on~regular expressions. \textsc{ASLA} shows a competitive performance~to~two~open source license analyzers, i.e., \textsc{OSLC}~\cite{OSLC} and \textsc{FOSSology}~\cite{fossology, gobeille2008fossology}.~However, they do not report unknown licenses whose regular expressions are not prepared. To address this problem, German et al.~\cite{german2010sentence} design \textsc{Ninka} to first break license statement into sentences~and~then match sentence-tokens with sentence-tokens already manually identified for each license (i.e., a license rule) to identify~licenses. \textsc{Ninka} will report an unknown license if no license is matched. To~help~generate license rules for licenses reported as unknown by \textsc{Ninka}, Higashi et al.~\cite{higashi2016clustering} leverage a hierarchical clustering to group unknown licenses into clusters of files with a single license. Besides,~Vendome et al.~\cite{vendome2017machine} propose a machine learning-based approach to identify~exceptions appended to licenses. Kapitsaki and Paschalides~\cite{kapitsaki2017identifying} design \textsc{FOSS-LTE} to identify license terms from license texts. \todo{Our~work~uses \textsc{Ninka} to identify declared licenses~in~each~file.}

\textbf{License Understanding.} Several studies have been conducted~to understand license usage and evolution in open source software~systems.~Manabe et al.~\cite{manabe2010evolutional} analyze license changes along with~the~evolution~of four open source systems. Similarly, Di Penta et~al.~\cite{di2010exploratory} automatically track license term changes in six open source systems. These two studies reveal that licenses change frequently~and~substantially. Vendome et al.~\cite{vendome2015license, vendome2017license} investigate when and why developers adopt and change licenses by analyzing commits and surveying developers. Almeida et al.~\cite{almeida2017software} explore whether developers understand the licenses they use through a survey. \todo{To the best~of~our knowledge, our work is the first to systematically characterize the modification terms in open source licenses.}


\textbf{License Bug Detection.} Vendome et al.~\cite{vendome2018distribute} conduct~a~study to characterize license bugs by building a catalog of license bugs~and~understanding their implications on the software projects they affect. A number of approaches have also been proposed~to~detect~license bugs (or license violations), and they often target two kinds~of~license bugs,~i.e.,~license~incompatibility and license inconsistency.

License incompatibility refers to the incompatibility among the licenses of a system and its declared dependencies~or~reused source code. German et al.~\cite{german2009license} propose a concept model to formally~specify licenses and detect incompatibilities among licenses of a system and its declared dependencies. They also manually identify license incompatibilities in eight systems, and summarize patterns to resolve license incompatibilities.~Mathur et al.~\cite{mathur2012empirical} and Golubev~et~al.~\cite{golubev2020study} empirically investigate~incompatibilities among licenses of a system and its reused~code via manual analysis. German and Di Penta~\cite{german2012method} propose~a~semiautomatic~approach \textsc{Kenen} to open source license compliance for~Java~systems.~However, it relies on manual analysis to detect license incompatibilities. Similarly, van der Burg~\cite{van2014tracing} propose~to~automatically~identify the~dependencies~of~a~system by~analyzing the build process, but manually identify license incompatibilities. Kapitsaki et al.~\cite{kapitsaki2017automating, paschalides2016validate}~model~license~compatibility into~a directed~graph,~which~automates~license~incompatibility detection with Software Package Data Exchange. Xu~et~al.~\cite{xu2023lidetector} leverage natural language processing to identify license terms and infer rights and obligations for incompatibility detection without~any~prior~knowledge about licenses. Hemel et al.~\cite{hemel2011finding}, Duan et al.~\cite{duan2017identifying} and Feng et al.~\cite{feng2019open} identify open source components used in a binary by binary analysis, and detect violations of GPL/AGPL licenses which require the binary to be open source if any open source component under the protection of GPL/AGPL is used in the binary.

License inconsistency refers to two cases, i.e, the inconsistency between the licenses of two source files that are evolved from the same provenance by code reuse, and the inconsistency between~the license of a package and the license of each file of the package. German et al.~\cite{german2009code} and Wu et al.~\cite{wu2015method, wu2017analysis} focus on the first~case.~They first use clone detection to find source files that have code clones, and use \textsc{Ninka} to identify licenses of these source files for inconsistency detection. German et al.~\cite{german2010understanding}, Di Penta et al.~\cite{di2010identifying} and~Mlouki~et~al.~\cite{mlouki2016detection} focus on the second case. They either use metadata or tools~like~\textsc{FOSSology} and \textsc{Ninka} to identify licenses for inconsistency detection.

To prevent such license bugs, Liu et al.~\cite{liu2019predicting} design~a~learning-based approach to predict~a compatible license when third-party~dependencies that cause~a~license incompatibility are imported. Kapitsaki and Charalambous~\cite{kapitsaki2019modeling} introduce a license recommender findOSSLicense to assist developers in choosing an appropriate open source license based on modeling license terms and utilizing knowledge from existing software projects.

\todo{Different from these studies, we focus on a new type~of~license bugs, i.e., violations of modification terms in open source licenses. To the best of our knowledge, we are the~first~to~detect and fix them for ensuring modification term~compliance.}

\textbf{Open~Source~Software~Reuse.}~Open~source~software~reuse is a common activity in software development in various~forms, e.g.,~copy and paste~\cite{kim2004ethnographic, kim2005empirical, juergens2009code}, forking~\cite{jiang2017and, zhou2020has, brisson2020we},~and~importing third-party dependencies~\cite{huang2022characterizing, zaimi2015empirical}. \todo{Our work is specifically focused on license bugs during forking. To the best~of~our knowledge, previous studies conduct empirical analysis~of~forking, but no work considers license bugs during forking.} For example, Jiang et al.~\cite{jiang2017and} explore developers' motivation and behavior on forking in GitHub. They~find that some developers fork repositories to add new features. These activities should be conducted under the license modification terms. Brisson et al.~\cite{brisson2020we} study the communication (e.g., pull requests, followers and contributors) of repositories which are from~the~same~software family via forking. Zhou et al.~\cite{zhou2020has} analyze the history~of forking in the past 20 years. They find that developers are less concerned about community~fragmentation~but perceive forks as~a~positive~noncompetitive~alternative~to~the original~project.




\section{Conclusions}

We have empirically investigated \modifyterm~in \todo{107} open source license. We find that \todo{47} licenses noticeably declare \modifyterm, and require certain forms~of ``notice'' to describe~modifications made to the original~work.~We~model the \modifyterm with respect to modification scope,~notice content and notice location. Further, we have designed \tool~to automatically detect and fix modification term violations during forking. Our evaluation results have demonstrated the severity of modification term violations in real-world fork repositories and the effectiveness and efficiency of \tool. We~have~release \tool's source code and experimental data at our website~\cite{zenodo}.


\bibliographystyle{ACM-Reference-Format}
\bibliography{src/reference}


\begin{thebibliography}{54}


\ifx \showCODEN    \undefined \def \showCODEN     #1{\unskip}     \fi
\ifx \showDOI      \undefined \def \showDOI       #1{#1}\fi
\ifx \showISBNx    \undefined \def \showISBNx     #1{\unskip}     \fi
\ifx \showISBNxiii \undefined \def \showISBNxiii  #1{\unskip}     \fi
\ifx \showISSN     \undefined \def \showISSN      #1{\unskip}     \fi
\ifx \showLCCN     \undefined \def \showLCCN      #1{\unskip}     \fi
\ifx \shownote     \undefined \def \shownote      #1{#1}          \fi
\ifx \showarticletitle \undefined \def \showarticletitle #1{#1}   \fi
\ifx \showURL      \undefined \def \showURL       {\relax}        \fi
\providecommand\bibfield[2]{#2}
\providecommand\bibinfo[2]{#2}
\providecommand\natexlab[1]{#1}
\providecommand\showeprint[2][]{arXiv:#2}

\bibitem[fil(2022)]%
        {filewithapache}
 \bibinfo{year}{2022}\natexlab{}.
\newblock \bibinfo{booktitle}{\emph{File With Apache 2.0 and My
  Modifications}}.
\newblock
\urldef\tempurl%
\url{https://softwareengineering.stackexchange.com/questions/220068/file-with-apache-2-0-and-my-modifications}
\showURL{%
Retrieved March 21, 2023 from \tempurl}


\bibitem[how(2022)]%
        {howtospecify}
 \bibinfo{year}{2022}\natexlab{}.
\newblock \bibinfo{booktitle}{\emph{How to Specify Notices of All Changes in a
  Class}}.
\newblock
\urldef\tempurl%
\url{https://opensource.stackexchange.com/questions/4419/aplv2-how-to-specify-notices-of-all-changes-in-a-class}
\showURL{%
Retrieved March 21, 2023 from \tempurl}


\bibitem[tmv(2022)]%
        {tmvdiscussion}
 \bibinfo{year}{2022}\natexlab{}.
\newblock \bibinfo{booktitle}{\emph{TVM Distribute Questions About Apache
  2.0}}.
\newblock
\urldef\tempurl%
\url{https://discuss.tvm.apache.org/t/tvm-distribute-questions-about-apache-2-0-license/8805/6}
\showURL{%
Retrieved March 21, 2023 from \tempurl}


\bibitem[Almeida et~al\mbox{.}(2017)]%
        {almeida2017software}
\bibfield{author}{\bibinfo{person}{D.~A Almeida}, \bibinfo{person}{G.~C
  Murphy}, \bibinfo{person}{G. Wilson}, {and} \bibinfo{person}{M. Hoye}.}
  \bibinfo{year}{2017}\natexlab{}.
\newblock \showarticletitle{Do software developers understand open source
  licenses?}. In \bibinfo{booktitle}{\emph{Proceedings of the IEEE/ACM 25th
  International Conference on Program Comprehension}}. \bibinfo{pages}{1--11}.
\newblock


\bibitem[Brisson et~al\mbox{.}(2020)]%
        {brisson2020we}
\bibfield{author}{\bibinfo{person}{S. Brisson}, \bibinfo{person}{E. Noei},
  {and} \bibinfo{person}{K. Lyons}.} \bibinfo{year}{2020}\natexlab{}.
\newblock \showarticletitle{We are family: analyzing communication in GitHub
  software repositories and their forks}. In
  \bibinfo{booktitle}{\emph{Proceedings of the IEEE 27th International
  Conference on Software Analysis, Evolution and Reengineering}}.
  \bibinfo{pages}{59--69}.
\newblock


\bibitem[daxlec et~al\mbox{.}(2007)]%
        {OSLC}
\bibfield{author}{\bibinfo{person}{daxlec}, \bibinfo{person}{devil\_moon},
  \bibinfo{person}{lrontyne}, \bibinfo{person}{sjkaaria}, {and}
  \bibinfo{person}{villoks}.} \bibinfo{year}{2007}\natexlab{}.
\newblock \bibinfo{booktitle}{\emph{OSLC}}.
\newblock
\urldef\tempurl%
\url{https://sourceforge.net/projects/oslc/}
\showURL{%
Retrieved March 21, 2023 from \tempurl}


\bibitem[Di~Penta et~al\mbox{.}(2010a)]%
        {di2010identifying}
\bibfield{author}{\bibinfo{person}{M. Di~Penta}, \bibinfo{person}{D.~M German},
  {and} \bibinfo{person}{G. Antoniol}.} \bibinfo{year}{2010}\natexlab{a}.
\newblock \showarticletitle{Identifying licensing of jar archives using a
  code-search approach}. In \bibinfo{booktitle}{\emph{Proceedings of the 7th
  IEEE Working Conference on Mining Software Repositories}}.
  \bibinfo{pages}{151--160}.
\newblock


\bibitem[Di~Penta et~al\mbox{.}(2010b)]%
        {di2010exploratory}
\bibfield{author}{\bibinfo{person}{Massimiliano Di~Penta},
  \bibinfo{person}{Daniel~M German}, \bibinfo{person}{Yann-Ga{\"e}l
  Gu{\'e}h{\'e}neuc}, {and} \bibinfo{person}{Giuliano Antoniol}.}
  \bibinfo{year}{2010}\natexlab{b}.
\newblock \showarticletitle{An exploratory study of the evolution of software
  licensing}. In \bibinfo{booktitle}{\emph{Proceedings of the 32nd ACM/IEEE
  International Conference on Software Engineering}}.
  \bibinfo{pages}{145--154}.
\newblock


\bibitem[Duan et~al\mbox{.}(2017)]%
        {duan2017identifying}
\bibfield{author}{\bibinfo{person}{R. Duan}, \bibinfo{person}{A. Bijlani},
  \bibinfo{person}{M. Xu}, \bibinfo{person}{T. Kim}, {and} \bibinfo{person}{W.
  Lee}.} \bibinfo{year}{2017}\natexlab{}.
\newblock \showarticletitle{Identifying open-source license violation and 1-day
  security risk at large scale}. In \bibinfo{booktitle}{\emph{Proceedings of
  the ACM SIGSAC Conference on computer and communications security}}.
  \bibinfo{pages}{2169--2185}.
\newblock


\bibitem[Eclipse(2023)]%
        {jgit}
\bibfield{author}{\bibinfo{person}{Eclipse}.} \bibinfo{year}{2023}\natexlab{}.
\newblock \bibinfo{booktitle}{\emph{JGit}}.
\newblock
\urldef\tempurl%
\url{https://www.eclipse.org/jgit/}
\showURL{%
Retrieved April 24, 2023 from \tempurl}


\bibitem[Feng et~al\mbox{.}(2019)]%
        {feng2019open}
\bibfield{author}{\bibinfo{person}{Muyue Feng}, \bibinfo{person}{Weixuan Mao},
  \bibinfo{person}{Zimu Yuan}, \bibinfo{person}{Yang Xiao}, \bibinfo{person}{Gu
  Ban}, \bibinfo{person}{Wei Wang}, \bibinfo{person}{Shiyang Wang},
  \bibinfo{person}{Qian Tang}, \bibinfo{person}{Jiahuan Xu},
  \bibinfo{person}{He Su}, {et~al\mbox{.}}} \bibinfo{year}{2019}\natexlab{}.
\newblock \showarticletitle{Open-source license violations of binary software
  at large scale}. In \bibinfo{booktitle}{\emph{Proceedings of the IEEE 26th
  International Conference on Software Analysis, Evolution and Reengineering}}.
  \bibinfo{pages}{564--568}.
\newblock


\bibitem[Foundation(2017)]%
        {fossology}
\bibfield{author}{\bibinfo{person}{Linux Foundation}.}
  \bibinfo{year}{2017}\natexlab{}.
\newblock \bibinfo{booktitle}{\emph{Fossology}}.
\newblock
\urldef\tempurl%
\url{https://www.fossology.org/}
\showURL{%
Retrieved March 21, 2023 from \tempurl}


\bibitem[German and Di~Penta(2012)]%
        {german2012method}
\bibfield{author}{\bibinfo{person}{Daniel German} {and}
  \bibinfo{person}{Massimiliano Di~Penta}.} \bibinfo{year}{2012}\natexlab{}.
\newblock \showarticletitle{A method for open source license compliance of java
  applications}.
\newblock \bibinfo{journal}{\emph{IEEE software}} \bibinfo{volume}{29},
  \bibinfo{number}{3} (\bibinfo{year}{2012}), \bibinfo{pages}{58--63}.
\newblock


\bibitem[German et~al\mbox{.}(2010a)]%
        {german2010understanding}
\bibfield{author}{\bibinfo{person}{D.~M German}, \bibinfo{person}{M. Di~Penta},
  {and} \bibinfo{person}{J. Davies}.} \bibinfo{year}{2010}\natexlab{a}.
\newblock \showarticletitle{Understanding and auditing the licensing of open
  source software distributions}. In \bibinfo{booktitle}{\emph{Proceedings of
  the IEEE 18th International Conference on Program Comprehension}}.
  \bibinfo{pages}{84--93}.
\newblock


\bibitem[German et~al\mbox{.}(2009)]%
        {german2009code}
\bibfield{author}{\bibinfo{person}{Daniel~M German},
  \bibinfo{person}{Massimiliano Di~Penta}, \bibinfo{person}{Yann-Gael
  Gueheneuc}, {and} \bibinfo{person}{Giuliano Antoniol}.}
  \bibinfo{year}{2009}\natexlab{}.
\newblock \showarticletitle{Code siblings: Technical and legal implications of
  copying code between applications}. In \bibinfo{booktitle}{\emph{Proceedings
  of the 6th IEEE International Working Conference on Mining Software
  Repositories}}. \bibinfo{pages}{81--90}.
\newblock


\bibitem[German and Hassan(2009)]%
        {german2009license}
\bibfield{author}{\bibinfo{person}{Daniel~M German} {and}
  \bibinfo{person}{Ahmed~E Hassan}.} \bibinfo{year}{2009}\natexlab{}.
\newblock \showarticletitle{License integration patterns: Addressing license
  mismatches in component-based development}. In
  \bibinfo{booktitle}{\emph{Proceedings of the IEEE 31st international
  conference on software engineering}}. \bibinfo{pages}{188--198}.
\newblock


\bibitem[German et~al\mbox{.}(2010b)]%
        {german2010sentence}
\bibfield{author}{\bibinfo{person}{D.~M German}, \bibinfo{person}{Y. Manabe},
  {and} \bibinfo{person}{K. Inoue}.} \bibinfo{year}{2010}\natexlab{b}.
\newblock \showarticletitle{A sentence-matching method for automatic license
  identification of source code files}. In
  \bibinfo{booktitle}{\emph{Proceedings of the IEEE/ACM international
  conference on Automated software engineering}}. \bibinfo{pages}{437--446}.
\newblock


\bibitem[Gobeille(2008)]%
        {gobeille2008fossology}
\bibfield{author}{\bibinfo{person}{Robert Gobeille}.}
  \bibinfo{year}{2008}\natexlab{}.
\newblock \showarticletitle{The fossology project}. In
  \bibinfo{booktitle}{\emph{Proceedings of the 2008 international working
  conference on Mining software repositories}}. \bibinfo{pages}{47--50}.
\newblock


\bibitem[Golubev et~al\mbox{.}(2020)]%
        {golubev2020study}
\bibfield{author}{\bibinfo{person}{Yaroslav Golubev}, \bibinfo{person}{Maria
  Eliseeva}, \bibinfo{person}{Nikita Povarov}, {and} \bibinfo{person}{Timofey
  Bryksin}.} \bibinfo{year}{2020}\natexlab{}.
\newblock \showarticletitle{A study of potential code borrowing and license
  violations in java projects on github}. In
  \bibinfo{booktitle}{\emph{Proceedings of the 17th International Conference on
  Mining Software Repositories}}. \bibinfo{pages}{54--64}.
\newblock


\bibitem[Grandini et~al\mbox{.}(2020)]%
        {grandini2020metrics}
\bibfield{author}{\bibinfo{person}{Margherita Grandini},
  \bibinfo{person}{Enrico Bagli}, {and} \bibinfo{person}{Giorgio Visani}.}
  \bibinfo{year}{2020}\natexlab{}.
\newblock \bibinfo{title}{Metrics for Multi-Class Classification: an Overview}.
\newblock
\newblock
\showeprint[arxiv]{2008.05756}~[stat.ML]


\bibitem[Hemel et~al\mbox{.}(2011)]%
        {hemel2011finding}
\bibfield{author}{\bibinfo{person}{A. Hemel}, \bibinfo{person}{K.~T.
  Kalleberg}, \bibinfo{person}{R. Vermaas}, {and} \bibinfo{person}{E.
  Dolstra}.} \bibinfo{year}{2011}\natexlab{}.
\newblock \showarticletitle{Finding software license violations through binary
  code clone detection}. In \bibinfo{booktitle}{\emph{Proceedings of the 8th
  Working Conference on Mining Software Repositories}}.
  \bibinfo{pages}{63--72}.
\newblock


\bibitem[Higashi et~al\mbox{.}(2016)]%
        {higashi2016clustering}
\bibfield{author}{\bibinfo{person}{Y. Higashi}, \bibinfo{person}{Y. Manabe},
  {and} \bibinfo{person}{M. Ohira}.} \bibinfo{year}{2016}\natexlab{}.
\newblock \showarticletitle{Clustering OSS license statements toward automatic
  generation of license rules}. In \bibinfo{booktitle}{\emph{Proceedings of the
  7th International Workshop on Empirical Software Engineering in Practice}}.
  \bibinfo{pages}{30--35}.
\newblock


\bibitem[Huang et~al\mbox{.}(2022)]%
        {huang2022characterizing}
\bibfield{author}{\bibinfo{person}{K. Huang}, \bibinfo{person}{B. Chen},
  \bibinfo{person}{C. Xu}, \bibinfo{person}{Y. Wang}, \bibinfo{person}{B. Shi},
  \bibinfo{person}{X. Peng}, \bibinfo{person}{Y. Wu}, {and} \bibinfo{person}{Y.
  Liu}.} \bibinfo{year}{2022}\natexlab{}.
\newblock \showarticletitle{Characterizing usages, updates and risks of
  third-party libraries in Java projects}.
\newblock \bibinfo{journal}{\emph{Empirical Software Engineering}}
  \bibinfo{volume}{27}, \bibinfo{number}{4} (\bibinfo{year}{2022}),
  \bibinfo{pages}{1--41}.
\newblock


\bibitem[Initiative(2022)]%
        {opensourcelic}
\bibfield{author}{\bibinfo{person}{Open~Source Initiative}.}
  \bibinfo{year}{2022}\natexlab{}.
\newblock \bibinfo{booktitle}{\emph{Open Source Licenses}}.
\newblock
\urldef\tempurl%
\url{https://opensource.org/licenses}
\showURL{%
Retrieved March 21, 2023 from \tempurl}


\bibitem[Jiang et~al\mbox{.}(2017)]%
        {jiang2017and}
\bibfield{author}{\bibinfo{person}{J. Jiang}, \bibinfo{person}{D. Lo},
  \bibinfo{person}{J. He}, \bibinfo{person}{X. Xia}, \bibinfo{person}{P.~S.
  Kochhar}, {and} \bibinfo{person}{L. Zhang}.} \bibinfo{year}{2017}\natexlab{}.
\newblock \showarticletitle{Why and how developers fork what from whom in
  GitHub}.
\newblock \bibinfo{journal}{\emph{Empirical Software Engineering}}
  \bibinfo{volume}{22}, \bibinfo{number}{1} (\bibinfo{year}{2017}),
  \bibinfo{pages}{547--578}.
\newblock


\bibitem[Jones(1972)]%
        {jones1972statistical}
\bibfield{author}{\bibinfo{person}{K. Jones}.} \bibinfo{year}{1972}\natexlab{}.
\newblock \showarticletitle{A statistical interpretation of term specificity
  and its application in retrieval}.
\newblock \bibinfo{journal}{\emph{Journal of documentation}}
  \bibinfo{volume}{28}, \bibinfo{number}{1} (\bibinfo{year}{1972}),
  \bibinfo{pages}{11--21}.
\newblock


\bibitem[Juergens et~al\mbox{.}(2009)]%
        {juergens2009code}
\bibfield{author}{\bibinfo{person}{E. Juergens}, \bibinfo{person}{F.
  Deissenboeck}, \bibinfo{person}{B. Hummel}, {and} \bibinfo{person}{S.
  Wagner}.} \bibinfo{year}{2009}\natexlab{}.
\newblock \showarticletitle{Do code clones matter?}. In
  \bibinfo{booktitle}{\emph{Proceedings of the IEEE 31st International
  Conference on Software Engineering}}. \bibinfo{pages}{485--495}.
\newblock


\bibitem[Kapitsaki and Charalambous(2019)]%
        {kapitsaki2019modeling}
\bibfield{author}{\bibinfo{person}{G.~M Kapitsaki} {and} \bibinfo{person}{G.
  Charalambous}.} \bibinfo{year}{2019}\natexlab{}.
\newblock \showarticletitle{Modeling and recommending open source licenses with
  findOSSLicense}.
\newblock \bibinfo{journal}{\emph{IEEE Transactions on Software Engineering}}
  \bibinfo{volume}{47}, \bibinfo{number}{5} (\bibinfo{year}{2019}),
  \bibinfo{pages}{919--935}.
\newblock


\bibitem[Kapitsaki et~al\mbox{.}(2017)]%
        {kapitsaki2017automating}
\bibfield{author}{\bibinfo{person}{G.~M Kapitsaki}, \bibinfo{person}{F.
  Kramer}, {and} \bibinfo{person}{N.~D Tselikas}.}
  \bibinfo{year}{2017}\natexlab{}.
\newblock \showarticletitle{Automating the license compatibility process in
  open source software with SPDX}.
\newblock \bibinfo{journal}{\emph{Journal of systems and software}}
  \bibinfo{volume}{131} (\bibinfo{year}{2017}), \bibinfo{pages}{386--401}.
\newblock


\bibitem[Kapitsaki and Paschalides(2017)]%
        {kapitsaki2017identifying}
\bibfield{author}{\bibinfo{person}{G.~M Kapitsaki} {and} \bibinfo{person}{D.
  Paschalides}.} \bibinfo{year}{2017}\natexlab{}.
\newblock \showarticletitle{Identifying terms in open source software license
  texts}. In \bibinfo{booktitle}{\emph{Proceedings of the 24th Asia-Pacific
  Software Engineering Conference}}. \bibinfo{pages}{540--545}.
\newblock


\bibitem[Kim et~al\mbox{.}(2004)]%
        {kim2004ethnographic}
\bibfield{author}{\bibinfo{person}{M. Kim}, \bibinfo{person}{L. Bergman},
  \bibinfo{person}{T. Lau}, {and} \bibinfo{person}{D. Notkin}.}
  \bibinfo{year}{2004}\natexlab{}.
\newblock \showarticletitle{An ethnographic study of copy and paste programming
  practices in OOPL}. In \bibinfo{booktitle}{\emph{Proceedings of the 2004
  International Symposium on Empirical Software Engineering}}.
  \bibinfo{pages}{83--92}.
\newblock


\bibitem[Kim et~al\mbox{.}(2005)]%
        {kim2005empirical}
\bibfield{author}{\bibinfo{person}{M. Kim}, \bibinfo{person}{V. Sazawal},
  \bibinfo{person}{D. Notkin}, {and} \bibinfo{person}{G. Murphy}.}
  \bibinfo{year}{2005}\natexlab{}.
\newblock \showarticletitle{An empirical study of code clone genealogies}. In
  \bibinfo{booktitle}{\emph{Proceedings of the 10th European software
  engineering conference held jointly with 13th ACM SIGSOFT international
  symposium on Foundations of software engineering}}.
  \bibinfo{pages}{187--196}.
\newblock


\bibitem[Li et~al\mbox{.}(2020)]%
        {li2020saga}
\bibfield{author}{\bibinfo{person}{G. Li}, \bibinfo{person}{Y. Wu},
  \bibinfo{person}{C.~K Roy}, \bibinfo{person}{J. Sun}, \bibinfo{person}{X.
  Peng}, \bibinfo{person}{N. Zhan}, \bibinfo{person}{B. Hu}, {and}
  \bibinfo{person}{J. Ma}.} \bibinfo{year}{2020}\natexlab{}.
\newblock \showarticletitle{SAGA: efficient and large-scale detection of
  near-miss clones with GPU acceleration}. In
  \bibinfo{booktitle}{\emph{Proceedings of the IEEE 27th International
  Conference on Software Analysis, Evolution and Reengineering}}.
  \bibinfo{pages}{272--283}.
\newblock


\bibitem[Liu et~al\mbox{.}(2019)]%
        {liu2019predicting}
\bibfield{author}{\bibinfo{person}{X. Liu}, \bibinfo{person}{L. Huang},
  \bibinfo{person}{J. Ge}, {and} \bibinfo{person}{V. Ng}.}
  \bibinfo{year}{2019}\natexlab{}.
\newblock \showarticletitle{Predicting licenses for changed source code}. In
  \bibinfo{booktitle}{\emph{Proceedings of the 34th IEEE/ACM International
  Conference on Automated Software Engineering}}. \bibinfo{pages}{686--697}.
\newblock


\bibitem[Luhn(1957)]%
        {luhn1957statistical}
\bibfield{author}{\bibinfo{person}{H. Luhn}.} \bibinfo{year}{1957}\natexlab{}.
\newblock \showarticletitle{A statistical approach to mechanized encoding and
  searching of literary information}.
\newblock \bibinfo{journal}{\emph{IBM Journal of research and development}}
  \bibinfo{volume}{1}, \bibinfo{number}{4} (\bibinfo{year}{1957}),
  \bibinfo{pages}{309--317}.
\newblock


\bibitem[Manabe et~al\mbox{.}(2010)]%
        {manabe2010evolutional}
\bibfield{author}{\bibinfo{person}{Yuki Manabe}, \bibinfo{person}{Yasuhiro
  Hayase}, {and} \bibinfo{person}{Katuro Inoue}.}
  \bibinfo{year}{2010}\natexlab{}.
\newblock \showarticletitle{Evolutional analysis of licenses in FOSS}. In
  \bibinfo{booktitle}{\emph{Proceedings of the Joint ERCIM Workshop on Software
  Evolution and International Workshop on Principles of Software Evolution}}.
  \bibinfo{pages}{83--87}.
\newblock


\bibitem[Mathur et~al\mbox{.}(2012)]%
        {mathur2012empirical}
\bibfield{author}{\bibinfo{person}{Arunesh Mathur}, \bibinfo{person}{Harshal
  Choudhary}, \bibinfo{person}{Priyank Vashist}, \bibinfo{person}{William
  Thies}, {and} \bibinfo{person}{Santhi Thilagam}.}
  \bibinfo{year}{2012}\natexlab{}.
\newblock \showarticletitle{An empirical study of license violations in open
  source projects}. In \bibinfo{booktitle}{\emph{Proceedings of the 35th Annual
  IEEE Software Engineering Workshop}}. \bibinfo{pages}{168--176}.
\newblock


\bibitem[Mlouki et~al\mbox{.}(2016)]%
        {mlouki2016detection}
\bibfield{author}{\bibinfo{person}{O. Mlouki}, \bibinfo{person}{F. Khomh},
  {and} \bibinfo{person}{G. Antoniol}.} \bibinfo{year}{2016}\natexlab{}.
\newblock \showarticletitle{On the detection of licenses violations in the
  android ecosystem}. In \bibinfo{booktitle}{\emph{Proceedings of the IEEE 23rd
  International Conference on Software Analysis, Evolution, and
  Reengineering}}, Vol.~\bibinfo{volume}{1}. \bibinfo{pages}{382--392}.
\newblock


\bibitem[Paschalides and Kapitsaki(2016)]%
        {paschalides2016validate}
\bibfield{author}{\bibinfo{person}{Demetris Paschalides} {and}
  \bibinfo{person}{Georgia~M Kapitsaki}.} \bibinfo{year}{2016}\natexlab{}.
\newblock \showarticletitle{Validate your SPDX files for open source license
  violations}. In \bibinfo{booktitle}{\emph{Proceedings of the 2016 24th ACM
  SIGSOFT International Symposium on Foundations of Software Engineering}}.
  \bibinfo{pages}{1047--1051}.
\newblock


\bibitem[Pedregosa et~al\mbox{.}(2011)]%
        {scikit-learn}
\bibfield{author}{\bibinfo{person}{F. Pedregosa}, \bibinfo{person}{G.
  Varoquaux}, \bibinfo{person}{A. Gramfort}, \bibinfo{person}{V. Michel},
  \bibinfo{person}{B. Thirion}, \bibinfo{person}{O. Grisel},
  \bibinfo{person}{M. Blondel}, \bibinfo{person}{P. Prettenhofer},
  \bibinfo{person}{R. Weiss}, \bibinfo{person}{V. Dubourg}, \bibinfo{person}{J.
  Vanderplas}, \bibinfo{person}{A. Passos}, \bibinfo{person}{D. Cournapeau},
  \bibinfo{person}{M. Brucher}, \bibinfo{person}{M. Perrot}, {and}
  \bibinfo{person}{E. Duchesnay}.} \bibinfo{year}{2011}\natexlab{}.
\newblock \showarticletitle{Scikit-learn: Machine Learning in {P}ython}.
\newblock \bibinfo{journal}{\emph{Journal of Machine Learning Research}}
  \bibinfo{volume}{12} (\bibinfo{year}{2011}), \bibinfo{pages}{2825--2830}.
\newblock


\bibitem[Seaman(1999)]%
        {seaman1999qualitative}
\bibfield{author}{\bibinfo{person}{Carolyn~B. Seaman}.}
  \bibinfo{year}{1999}\natexlab{}.
\newblock \showarticletitle{Qualitative methods in empirical studies of
  software engineering}.
\newblock \bibinfo{journal}{\emph{IEEE Transactions on software engineering}}
  \bibinfo{volume}{25}, \bibinfo{number}{4} (\bibinfo{year}{1999}),
  \bibinfo{pages}{557--572}.
\newblock


\bibitem[\tool(2023)]%
        {zenodo}
\bibfield{author}{\bibinfo{person}{\tool}.} \bibinfo{year}{2023}\natexlab{}.
\newblock \bibinfo{booktitle}{\emph{The Replication Package of \tool}}.
\newblock
\urldef\tempurl%
\url{https://doi.org/10.5281/zenodo.7902495}
\showURL{%
Retrieved March 21, 2023 from \tempurl}


\bibitem[Tuunanen et~al\mbox{.}(2009)]%
        {tuunanen2009automated}
\bibfield{author}{\bibinfo{person}{T. Tuunanen}, \bibinfo{person}{J. Koskinen},
  {and} \bibinfo{person}{T. K{\"a}rkk{\"a}inen}.}
  \bibinfo{year}{2009}\natexlab{}.
\newblock \showarticletitle{Automated software license analysis}.
\newblock \bibinfo{journal}{\emph{Automated Software Engineering}}
  \bibinfo{volume}{16}, \bibinfo{number}{3} (\bibinfo{year}{2009}),
  \bibinfo{pages}{455--490}.
\newblock


\bibitem[Van Der~Burg et~al\mbox{.}(2014)]%
        {van2014tracing}
\bibfield{author}{\bibinfo{person}{S. Van Der~Burg}, \bibinfo{person}{E.
  Dolstra}, \bibinfo{person}{S. McIntosh}, \bibinfo{person}{J. Davies},
  \bibinfo{person}{D.~M German}, {and} \bibinfo{person}{A. Hemel}.}
  \bibinfo{year}{2014}\natexlab{}.
\newblock \showarticletitle{Tracing software build processes to uncover license
  compliance inconsistencies}. In \bibinfo{booktitle}{\emph{Proceedings of the
  29th ACM/IEEE international conference on Automated software engineering}}.
  \bibinfo{pages}{731--742}.
\newblock


\bibitem[Vendome et~al\mbox{.}(2017a)]%
        {vendome2017license}
\bibfield{author}{\bibinfo{person}{Christopher Vendome},
  \bibinfo{person}{Gabriele Bavota}, \bibinfo{person}{Massimiliano~Di Penta},
  \bibinfo{person}{Mario Linares-V{\'a}squez}, \bibinfo{person}{Daniel German},
  {and} \bibinfo{person}{Denys Poshyvanyk}.} \bibinfo{year}{2017}\natexlab{a}.
\newblock \showarticletitle{License usage and changes: a large-scale study on
  github}.
\newblock \bibinfo{journal}{\emph{Empirical Software Engineering}}
  \bibinfo{volume}{22} (\bibinfo{year}{2017}), \bibinfo{pages}{1537--1577}.
\newblock


\bibitem[Vendome et~al\mbox{.}(2018)]%
        {vendome2018distribute}
\bibfield{author}{\bibinfo{person}{C. Vendome}, \bibinfo{person}{D. German},
  \bibinfo{person}{M. Di~Penta}, \bibinfo{person}{G. Bavota},
  \bibinfo{person}{M. Linares-V{\'a}squez}, {and} \bibinfo{person}{D.
  Poshyvanyk}.} \bibinfo{year}{2018}\natexlab{}.
\newblock \showarticletitle{To distribute or not to distribute? why licensing
  bugs matter}. In \bibinfo{booktitle}{\emph{Proceedings of the IEEE/ACM 40th
  International Conference on Software Engineering}}.
  \bibinfo{pages}{268--279}.
\newblock


\bibitem[Vendome et~al\mbox{.}(2015)]%
        {vendome2015license}
\bibfield{author}{\bibinfo{person}{C. Vendome}, \bibinfo{person}{M
  Linares-V{\'a}squez}, \bibinfo{person}{G. Bavota}, \bibinfo{person}{M.
  Di~Penta}, \bibinfo{person}{D. German}, {and} \bibinfo{person}{D.
  Poshyvanyk}.} \bibinfo{year}{2015}\natexlab{}.
\newblock \showarticletitle{License usage and changes: a large-scale study of
  java projects on github}. In \bibinfo{booktitle}{\emph{Proceedings of the
  IEEE 23rd International Conference on Program Comprehension}}.
  \bibinfo{pages}{218--228}.
\newblock


\bibitem[Vendome et~al\mbox{.}(2017b)]%
        {vendome2017machine}
\bibfield{author}{\bibinfo{person}{C. Vendome}, \bibinfo{person}{M.
  Linares-V{\'a}squez}, \bibinfo{person}{G. Bavota}, \bibinfo{person}{M.
  Di~Penta}, \bibinfo{person}{D. German}, {and} \bibinfo{person}{D.
  Poshyvanyk}.} \bibinfo{year}{2017}\natexlab{b}.
\newblock \showarticletitle{Machine learning-based detection of open source
  license exceptions}. In \bibinfo{booktitle}{\emph{Proceedings of the IEEE/ACM
  39th International Conference on Software Engineering}}.
  \bibinfo{pages}{118--129}.
\newblock


\bibitem[wikipedia(2023)]%
        {unlicense}
\bibfield{author}{\bibinfo{person}{wikipedia}.}
  \bibinfo{year}{2023}\natexlab{}.
\newblock \bibinfo{booktitle}{\emph{Unlicense}}.
\newblock
\urldef\tempurl%
\url{https://en.wikipedia.org/wiki/Unlicense}
\showURL{%
Retrieved March 21, 2023 from \tempurl}


\bibitem[Wu et~al\mbox{.}(2015)]%
        {wu2015method}
\bibfield{author}{\bibinfo{person}{Y. Wu}, \bibinfo{person}{Y. Manabe},
  \bibinfo{person}{T. Kanda}, \bibinfo{person}{D.~M German}, {and}
  \bibinfo{person}{K. Inoue}.} \bibinfo{year}{2015}\natexlab{}.
\newblock \showarticletitle{A method to detect license inconsistencies in
  large-scale open source projects}. In \bibinfo{booktitle}{\emph{Proceedings
  of the IEEE/ACM 12th Working Conference on Mining Software Repositories}}.
  \bibinfo{pages}{324--333}.
\newblock


\bibitem[Wu et~al\mbox{.}(2017)]%
        {wu2017analysis}
\bibfield{author}{\bibinfo{person}{Y. Wu}, \bibinfo{person}{Y. Manabe},
  \bibinfo{person}{T. Kanda}, \bibinfo{person}{D.~M German}, {and}
  \bibinfo{person}{K. Inoue}.} \bibinfo{year}{2017}\natexlab{}.
\newblock \showarticletitle{Analysis of license inconsistency in large
  collections of open source projects}.
\newblock \bibinfo{journal}{\emph{Empirical Software Engineering}}
  \bibinfo{volume}{22}, \bibinfo{number}{3} (\bibinfo{year}{2017}),
  \bibinfo{pages}{1194--1222}.
\newblock


\bibitem[Xu et~al\mbox{.}(2023)]%
        {xu2023lidetector}
\bibfield{author}{\bibinfo{person}{Sihan Xu}, \bibinfo{person}{Ya Gao},
  \bibinfo{person}{Lingling Fan}, \bibinfo{person}{Zheli Liu},
  \bibinfo{person}{Yang Liu}, {and} \bibinfo{person}{Hua Ji}.}
  \bibinfo{year}{2023}\natexlab{}.
\newblock \showarticletitle{LiDetector: License Incompatibility Detection for
  Open Source Software}.
\newblock \bibinfo{journal}{\emph{ACM Transactions on Software Engineering and
  Methodology}} \bibinfo{volume}{32}, \bibinfo{number}{1}
  (\bibinfo{year}{2023}), \bibinfo{pages}{1--28}.
\newblock


\bibitem[Zaimi et~al\mbox{.}(2015)]%
        {zaimi2015empirical}
\bibfield{author}{\bibinfo{person}{A. Zaimi}, \bibinfo{person}{A. Ampatzoglou},
  \bibinfo{person}{N. Triantafyllidou}, \bibinfo{person}{A. Chatzigeorgiou},
  \bibinfo{person}{A. Mavridis}, \bibinfo{person}{T. Chaikalis},
  \bibinfo{person}{I. Deligiannis}, \bibinfo{person}{P. Sfetsos}, {and}
  \bibinfo{person}{I. Stamelos}.} \bibinfo{year}{2015}\natexlab{}.
\newblock \showarticletitle{An empirical study on the reuse of third-party
  libraries in open-source software development}. In
  \bibinfo{booktitle}{\emph{Proceedings of the 7th Balkan Conference on
  Informatics Conference}}. \bibinfo{pages}{1--8}.
\newblock


\bibitem[Zhou et~al\mbox{.}(2020)]%
        {zhou2020has}
\bibfield{author}{\bibinfo{person}{S. Zhou}, \bibinfo{person}{B. Vasilescu},
  {and} \bibinfo{person}{C. K{\"a}stner}.} \bibinfo{year}{2020}\natexlab{}.
\newblock \showarticletitle{How has forking changed in the last 20 years? a
  study of hard forks on github}. In \bibinfo{booktitle}{\emph{Proceedings of
  the IEEE/ACM 42nd International Conference on Software Engineering}}.
  \bibinfo{pages}{445--456}.
\newblock


\end{thebibliography}

\end{document}